\documentclass[apj]{emulateapj}

\shorttitle{X-ray Variability of AGN}
\shortauthors{Allevato et al.}
\begin{document}

\slugcomment{Accepted for publication in The Astrophysical Journal}

\title{Measuring X-ray variability in faint/sparsely-sampled AGN}

\author{V. Allevato\altaffilmark{1,2}, 
M. Paolillo\altaffilmark{2,3},
I. Papadakis\altaffilmark{4,5},
C. Pinto\altaffilmark{6}}

\altaffiltext{1}{Department of Physics, University of Helsinki, Gustaf H\"allstr\"omin katu 2a, FI-00014 Helsinki, Finland}
\altaffiltext{2}{Department of Physical Sciences, University Federico II, via Cinthia 6, 80126 Naples, Italy}
\altaffiltext{3}{Istituto Nazionale di Fisica Nucleare, Sez.di Napoli, Italy}
\altaffiltext{4}{Department of Physics and Institute of Theoretical \& Computational Physics, University of Crete, 71003, Heraklion, Greece}
\altaffiltext{5}{IESL, Foundation for Research and Technology, 71110, Heraklion, Greece}
\altaffiltext{6}{Institute of Astronomy, University of Cambridge, Madingley Road, CB3 0HA, Cambridge}

\def\sNXV{$\sigma^2_{\rm NXV}$}
\def\sML{$\sigma^2_{\rm ML}$}
\def\sbn{$\sigma^2_{\rm band,norm}$}
\def\scon{$\sigma^2_{\rm NXV,cont}$}
\def\sspa{$\sigma^2_{\rm NXV,sparse}$}

\begin{abstract}

We study the statistical properties of the \emph{Normalized Excess Variance} of
variability process characterized by a ``red-noise'' power spectral
 density (PSD), as the case of Active Galactic Nuclei (AGN). 
      We perform Monte Carlo simulations of lightcurves,
      assuming both a continuous and a sparse sampling pattern and
      various signal-to-noise (S/N) ratios. We show that the normalized
      excess variance is a biased estimate of the variance even in the
      case of continuously sampled lightcurves. The bias
      depends on the PSD slope and on the sampling pattern, but not on
      the S/N ratio. We provide a simple formula to
      account for the bias, which yields unbiased estimates with an accuracy
      better than $15$\%. We show that the normalized excess variance
      estimates based on single lightcurves (especially for sparse
      sampling and S/N $<$ 3) are highly uncertain (even if corrected
      for bias) and we propose instead the use of an ``ensemble estimate'', 
      based on multiple lightcurves of the same object, or on the use of lightcurves of
      many objects. These estimates have symmetric distributions, known
      errors, and can also be corrected for biases.We use our
      results to estimate the ability to measure the intrinsic source
      variability in current data, and show that they could also be
      useful in the planning of the observing strategy of future surveys
      such as those provided by X-ray missions studying distant and/or
      faint AGN populations and, more in general, in the estimation of
      the variability amplitude of sources that will result from future
      surveys such as Pan-STARRS, and LSST.

\end{abstract}


\section{Introduction}
\label{sec:intro}

Active Galactic Nuclei (AGN) are characterized by large amplitude and rapid
variability, especially in the X-ray band, which is probably originating in the
inner regions of the accretion disk and the hot corona in Seyfert AGNs. 
One of the most common
tools for examining AGN variability is the \emph{Power Spectral Density
Function} (PSD).  Early attempts to measure the AGN X-ray PSDs showed that they
have a power-law like shape  with a slope of $\sim -1.5$ \citep{Green93,
Lawrence93}, although PSDs as steep as $-3$ have been observed 
in X-ray light curves of radio-loud sources \citep{Kata01} or in optical AGN lightcurves \citep{Mush11}. 
This result is indicative of  a scale-invariant \emph{red-noise}
process, on timescales ranging from a few hours to years, with no evidence of
periodicities (with the possible exception of RE J1034+396 reported by \citealp{Gie08}).

In recent years it has become increasingly clear that there exists at least one
characteristic timescale in the AGN X-ray PSDs. This timescale reveals itself in
the form of "frequency breaks" ($\nu_{\rm br}$)  in the PSD, where the slope
changes from a value of $\sim -1$ below the  "break", to $\sim -2$ at
frequencies higher than $\nu_{\rm br}$ (see e.g. \citealt{Uttley02,
Markowitz03}). These time scales may be linked to the
characteristic disk time scales like the dynamical, thermal or viscous
timescale, and appear to correlate with the black hole (BH) mass and accretion rate
\citep{McHardy06, Koerding07, Gonz12}. Thus variability measurements represent a tool to
investigate both the physics of the accretion process, as well as the
fundamental parameters ($M_{\rm BH}$, $\dot{m}$) of the active nucleus.

So far, our knowledge of the X-ray variability properties of AGN is mainly
based on the study of a few nearby, X-ray bright objects, which have been
monitored extensively with Rossi X-ray Timing Explorer (\textit{RXTE}) over many years, and for which there also
exist day-long, high signal-to-noise (S/N)  {\it {\it XMM-Newton}} lightcurves. 
However several authors have also studied the ensemble variability properties of AGN
populations making use of different statistics 
\citep[e.g.,][]{ONeill05, Zhou10, Vagnetti11, Zhang11, Ponti12}.
The most commonly adopted statistics in these studies 
for the measure of the variability amplitude is the so called \textit{normalized excess  variance}, \sNXV\ 
\citep{Nandra97,Turner1999,Edelson2002}.

One of the main results of these investigations is that \sNXV\
anti-correlates with the mass of the central compact object 
\citep[e.g.,][]{Papadakis04,NikPapCzer04,ONeill05,Zhou10}. 
In fact, it has been suggested that \sNXV\ can be used to measure the mass of the central black
hole (BH) in AGN \citep[e.g.,][]{NikPapCzer04,GieNikCezr08, Ponti12}
and in other objects like the 
Ultraluminous X-ray sources \citet{Gonz11}. 
Furthermore, recent deep multi-cycle surveys (e.g. \citealt{Alexander2003,
Brunner2008, Comastri2011, Xue2011};  also see \citealt{Brandt2005} and
references therein), have been accumulating observations of intermediate and
high ($z>0.5$) redshift AGN, thus offering the opportunity to explore AGN
variability at high redshift as well. However, due to the sparse (i.e. uneven and with large gaps) sampling,
and the low flux of most AGN detected in these surveys, it is not possible to use
PSD techniques to study the variability properties of these objects.  Also in
these cases \sNXV\ has been used to parametrize the
variability properties of the high redshift AGN \citep{Almaini, Paolillo04,
Papadakis08}.

Despite the relative importance of the normalized excess  variance as a tool to
measure variability amplitude in AGN, as well as a tool to measure BH mass in
these, and perhaps other accreting objects as well, 
there have not been many studies to investigate its statistical properties. 
A systematic discussion of the statistical properties of \sNXV\ 
and its performance in the case of red noise PSDs of various slopes
and different signal-to-noise (S/N) ratios, can be found in
\citet{Vaughan03}. This work demonstrated that \sNXV\ 
extrapolations from any single realization can be misleading due to the stochastic 
nature of any red-noise lightcurve, and quantifies the expected uncertainties 
in order to, e.g. compare the same lightcurve in different energy bands.
However, even these authors have not considered explicitly
the question whether \sNXV\ is an unbiased estimate of the intrinsic
source variance or not, and have considered the case of  continuously sampled data
only, such as those provided by long XMM observations of nearby AGNs. Instead,
in serendipitous datasets, as well as in deep multi-cycle surveys, the
effects of sparse sampling must be taken taken into account when
investigating the statistical properties of the excess variance.

To some extent, this work is thus an extension of the work done by Vaughan et al (2003). 
Our goal is to investigate the statistical properties  of the excess variance 
(i.e. scatter, and the {\it mean}) in the case of both evenly {\it and} sparsely sampled light curves, 
whose PSD has a ``red noise" shape. We pay particular attention to the case of highly 
uneven patterns, like those in light curves which result from current multi-epoch surveys. 
This pattern is characterized by extreme sparsity due to the observing strategy and orbital visibility of the
targets. We consider various PSD slopes, sampling patterns, as well 
as S/N ratios, and we perform detailed Monte-Carlo numerical experiments to 
investigate the statistical properties (mean, standard deviation and skewness) 
of the excess variance in each case. The results are used to derive some simple recipes to
acquire excess variance measurements that will be unbiased estimates of the intrinsic variance (to a large extent), 
and will follow a Gaussian distribution with known errors, thus rendering them eligible to compare with 
theoretical predictions using the frequently used $\chi^2$ minimization techniques. 

\section{Normalized Excess Variance}
\label{sec:2}

The  so called \textit{normalized excess variance}, $\sigma^2_{\rm NXV}$,  is defined as \citep{Nandra97}:

 \begin{equation}
  \sigma^2_{\rm NXV} = \frac{1}{N \overline{x}^2} \sum_{i=1}^{N} \,
                   [(x_i - \overline{x})^2 - \sigma_{\rm err,\textit{i}}^2],
  \label{eq:exvar2}
\end{equation}

where $x_i$ and $\sigma_{\rm err,\textit{i}}$ are the count rate and its error in i-th bin, 
$\overline{x}$ is the mean count rate , and $N$ is the number of bins used to estimate 
\sNXV. With this normalization we are able to compare excess variance estimates 
derived from different segments of a particular lightcurve or from lightcurves of different sources. 
\citet{Nandra97} also provided an error estimate\footnote{There was a 
typographical error in \citet{Nandra97}, in that the equation for the error on \sNXV\ 
should have had the quantity inside the rms summation squared, as clarified by 
\cite{Turner1999}.} on \sNXV, which is based on the variance of the quantity $(x_i - \overline{x})^2 - \sigma_{\rm err,\textit{i}}^2$, i.e.
\begin{eqnarray}
  \Delta \sigma^2_{\rm NXV} & = & S_D / [\overline{x}^2 (N)^{1/2}],   \label{eq:exvar_err}       \\
  S_D & = & \frac{1}{N-1} \sum_{i=1}^{N} \,
                   \{[(x_i - \overline{x})^2 - \sigma_{\rm err,\textit{i}}^2]-
                   \sigma^2_{\rm NXV} \overline{x}^2\}^2.
                   \nonumber
  \end{eqnarray}

Almaini et al (2000) proposed an alternative normalized excess variance estimate, \sML, 
which they argued will perform better in the case when the errors on the lightcurve 
points are not identical and they are not normally distributed. There is no analytical 
equation for this estimate, as it is based on a maximum-likelihood approach, and 
the estimate has to be determined numerically for a given lightcurve (see their \S\ 3.1 for details).

If $\mu$ and $\sigma^2$ are the intrinsic mean and variance of a lightcurve, 
\sNXV\ and \sML\ are though to be an estimate of intrinsic normalized source 
variance, i.e.  $\sigma^2_{\rm norm}=\sigma^2/\mu^2$. However, this assumption 
has never been investigated thoroughly in the case when the light curve in 
question is a realization of process which has an intrinsic "red-noise" power 
spectrum. In this case, there are a few reasons to believe that this assumption 
may not hold. This can be understood if one considers the fact that the intrinsic 
power spectral density function, $PSD(\nu)$, of a time series is defined in such a way so that: 

\begin{equation}
\label{eq:sigmain}
\sigma^2=\int_{0}^{\infty} PSD(\nu) d\nu.
\end{equation}

As we mentioned above, based on the detailed studies of $\sim$15-20 nearby AGN, 
these sources (as well as the Galactic X-ray accreting objects) exhibit a power-law 
X-ray PSD at high frequencies of the form of $PSD(\nu) \propto \nu^{- \beta}$, 
with $\beta \sim 2$. There exists a "break frequency", $\nu_{\rm br}$, where the 
PSD slope flattens to a slope of $\beta \sim 1$ at lower frequencies. 
This "break frequency" depends on the BH mass of the system, and the respective "break 
time scale" (i.e. $1/\nu_{\rm br}$) increases proportionally with the BH mass 
\citep[e.g.,][]{McHardy06,Gonz12}.
It is of the order of $\sim$ few tens of minutes for 
BH masses of $\sim 10^6$ M$_{\odot}$, and increases to $\sim$ a 
day or so for BH masses $\sim 10^8$ M$_{\odot}$.  A second break to a 
slope of  $\sim 0$ is expected at even lower frequencies, for the integral in 
Eq. \ref{eq:sigmain} to converge, as expected for a stationary process. 
Such breaks are routinely observed in Galactic X-ray Black Hole binary candidates
when in the so called "low/hard" state. It has also been observed in one AGN 
(namely Ark 564, \citealt{Papadakis02,McHardy07}). However, in all other cases, 
the AGN PSDs exhibit a power-law shape with a 
slope of $\beta \sim 1$ at all sampled frequencies, which can 
be as low as $1/(10-20)$ years$^{-1}$ in the case of a few AGN 
which were regularly monitored by {\it RXTE} for more than a decade 
(e.g. NGC 4051, \citealt{McHardy04}; MCG--6-30-15, \citealt{McHardy05}). 

If \sNXV\ or \sML\ are computed using a lightcurve whose duration, 
$T_{\rm max}$, is shorter than the time scale at which the transition from 
the slope of $-1$ to zero appears in the intrinsic PSD, then: a) 
$\overline{x}$ may not be an accurate estimate of $\mu$,  so 
one has to investigate what are the effects of using $\overline{x}$ 
(as opposed to $\mu$) in the definition of \sNXV\ in 
Eq. \ref{eq:exvar2}, and b) \sNXV\ or \sML\ may {\it underestimate} 
$\sigma^2_{\rm norm}$, as there will be intrinsic variations on time 
scales longer than $T_{\rm max}$, which cannot be fully sampled in the given light curve. 

For these reasons \citet{Ponti12} as well as most of the aforementioned papers 
by Nikolajuk et al., Gonzalez et al., O'Neill et al. and \citet{Papadakis08} have assumed that \sNXV\ is 
a measure of the intrinsic "band" normalized variance, defined as, 

\begin{equation}
\label{eq:sigmabn}
\sigma^2_{\rm band,norm}= \left[  \int_{1/T_{\rm max}}^{1/T_{\rm min}} PSD(\nu) d\nu \right] /\mu^2.
\end{equation}
where $\mu$ and $PSD(\nu)$ are the intrinsic mean and PSD of the time series.

The "band" normalized variance measures the  
contribution to the total variance (normalized to the mean squared) 
of all the variability components 
with frequencies higher than 1/$T_{\rm max}$ (i.e. the longest, fully sampled 
frequency) and lower than 1/$T_{\rm min}$ (i.e. the highest sample frequency
since $T_{\rm min}=2\times \Delta t$, where $\Delta t$ is the the bin size of the observed light curve). 

However, even the assumption that \sNXV\ or \sML\ are estimates of the  
\sbn\ has not been tested in practice, and there are reasons to believe that 
it may not be entirely accurate in the case of "red-noise'' PSDs. The first reason is that, although 
variations on time scales longer than $T_{\rm max}$ are not fully sampled 
in a lightcurve with a length of $T_{\rm max}$, they can still contribute 
to its variance (this is the so-called "red-noise leakage" problem). 
As a result, \sNXV\ may {\it overestimate} \sbn. 
Secondly, in the case of sparsely sampled lightcurves, not all 
variability components will be sampled with the same "accuracy'', while the 
ability to recover intrinsic \sbn\ for faint sources may be undermined 
by the relatively stronger experimental noise level. 
Finally, in reality, the intrinsic PSD should continue with a power-law shape to frequencies higher than $1/T_{\rm min}$. Consequently, the observed normalized excess variance may again {\it overestimate} the intrinsic ``band" normalised variance, as defined in equation (4), due to aliasing effects. However, since the observed light curves are not sampled every $\Delta t$, but they are {\it binned} over intervals of size equal to $\Delta t$, any variations at frequencies higher than $1/(2\Delta t)$ are smeared out, and we do not expect the estimated normalized excess variance to be significantly affected by the intrinsic power at these high frequencies.

In the following sections we perform detailed Monte-Carlo simulations 
with the intent to verify if  \sNXV\ and \sML\ can be considered as accurate estimates of  
\sbn\ in the case of red-noise PSDs, and we further explore their 
statistical properties in the case of sparse sampling and sources with low S/N ratio lightcurves. 

\section{Monte Carlo Simulations: The algorithm and the relevant statistical parameters}
\label{sec:3}

\label{section:AGN_lc}

\begin{figure*}
\begin{center}
\includegraphics[width=18cm]{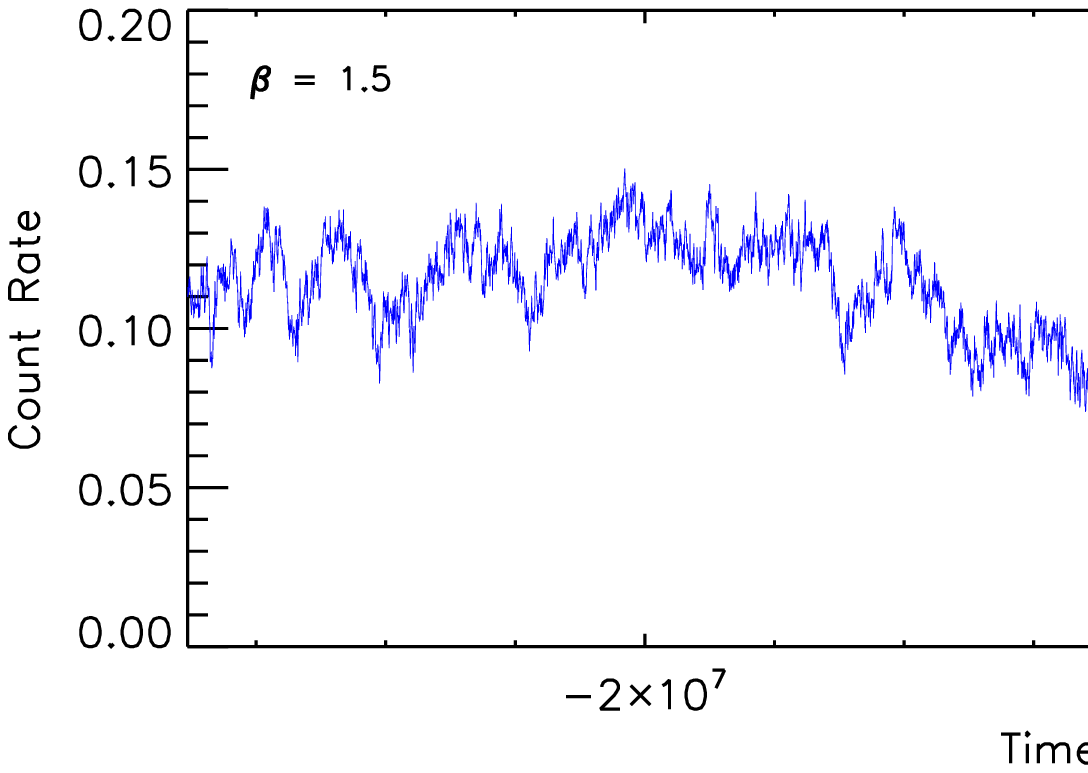}
        \caption{ 
        Simulated lightcurve (continuous blue line) 
        in the case of a PSD with $\beta=1.5$. Black 
        crosses with errorbars show the "continuously sampled lightcurve''  while the red dots indicate the "sparsely sampled lightcurve''. }
\label{fig:lc_sparsyXMM}
\end{center}
\end{figure*}

We performed Monte Carlo simulations by using the code of \citet{Timmer95}. 
This code is appropriate for the generation of Gaussian light curves with a given 
PSD. It is not clear whether Gaussianity applies to the AGN light curves (in any spectral band), and in fact, it is almost certain that it does not apply to the light curves of blazars. However, exploring the effects of non-Gaussianity to our results are beyond the scope of our paper. We also assume an AGN PSDs 
of light curves that are Gaussian and that remain only
weakly non-stationary (following the definition as in Section 4.1 of \citealt{Vaughan03})
on time scales of decades and shorter.
As for Gaussianity, 
exploring the effects of non-stationarity is beyond the scope of this work.
We modified the original code that generates red-noise data 
with a power law PSD, in order to reproduce the real data extraction 
process including filtering and background subtraction. 
We first created an evenly sampled AGN lightcurve with the above algorithm, 
following the appropriate PSD (see below). 
Then we added to the AGN count rate, 
in each time bin, the contribution from the expected (instrumental and cosmic) background
along the line-of-sight to the source, 
randomly adding Poisson fluctuations to both terms. A second 
 background estimate is also generated
(including again Poisson fluctuations), simulating the
one measured in a nearby region in real data, and then subtracted from 
the total counts in each bin, as done for real sources.

As a starting point, we simulated red-noise lightcurves with intrinsic 
count rate of $\mu_{\rm full}=0.099$ cnt s$^{-1}$ and variance, $\sigma^2_{\rm full}$,
such that $\sigma^2_{\rm full,norm}=\sigma^2_{\rm full}/\mu_{\rm full}^2=0.042 $ (see Table \ref{tab:input_param}). 
These are equal to the mean and normalized variance (after correcting for the Poisson noise) 
of the brightest AGN observed by {\it XMM-Newton}
in the CDFS \citep[source id 68 from][]{Gia02} at $z\sim 0.54$.
The source has an identical soft 0.5-2 keV and hard 2-8 keV flux of $\sim 5 \times 10^{-14}$ erg cm$^{-2}$ s$^{-1}$, i.e. 
we expect 10-20 of these sources per square degree, according to, e.g. 
\cite{Hasinger93,Luo08}.  Compared to other bright AGN in the field, this 
source has the advantage of being fairly isolated and thus its flux and
variability can be robustly estimated. 
We point out that in our simulations we assume an PSD with the 
appropriate normalization in order to yield the required \textit{intrinsic} variance and flux,
and we do not renormalize the lightcurves \textit{a-posteriori} after creating them\footnote{As done by 
some versions of the \citet{Timmer95} code found online.}, since the latter 
method would produce lightcurves which do not span the full range of
mean fluxes and variances.

We considered the sampling pattern of the first 1Ms 
{\it XMM-Newton} observations of the {\it Chandra Deep Field South} (CDFS, \citealp{Comastri2011}).  
{\it XMM-Newton} observed CDFS once in July 2001 with an effective exposure, 
after filtering high background periods, of 
$\sim 80$ ks and then, six times in January 2002 for additional $370$ ks. 
The time interval between the start of the first observation and the end of the last one (i.e. the longest time scale sampled by the 1Ms XMM-CDFS 
lightcurves) is $T_{\rm max}=1.56\times 10^7 $sec, i.e. about 6 months.
The actual XMM-CDFS observations lasted for $\sim 4.5\times 10^5$ 
sec ($\sim$ 5 days) during this period of 6 months. This type of observing 
pattern is a "worst-case" scenario (in terms of sparse sampling): the total exposure time is only $\sim 3$\% of the observing 
period, and the data are collected in two blocks at the start and at the end of the observing period.
This type of observing pattern is driven primarily by the typical scheduling requirements of deep 
multi-cycle campaigns, and thus represents a recurring, although undesirable, observing scheme 
\footnote{Only in 2009 an extended XMM-CDFS campaign allowed to mitigate the sampling problem, see 
\citet{Comastri2011}}. We nevertheless adopted it in order to study the effects of sparse 
sampling in an "extreme" case,  before discussing, 
in the next sections, more favourable sampling patterns.

We considered lightcurves whose PSD has a power-law shape of the form: 
$PSD(\nu)\propto \nu^{-\beta}$, where $\beta=1, 1.5, 2, 2.5 $ and 3.  
In the figures throughout the paper we present 
the results  for the simulations with $\beta=1.5$,
but we always discuss the results in the other cases as well. 
In order to account for the effect of red noise leak, which transfers 
power from low (undersampled) to high frequencies, we generated lightcurves which 
are 5 times longer than the longest timescale sampled by the data we considered.
We verified in the case of $\beta=3$ that extending the simulated lightcurves  by a factor of 10 does not significantly 
changes our results, while increasing considerably the processing time. 
Hence, we are confident that our simulations take into account properly 
the "red-noise" leak effects, even in the case of the PSDs with slopes steeper than 2. 

For each $\beta$, we produced 5000 simulated lightcurves with a 
length equal to $T_{\rm max,full}=5 \times T_{\rm max}$. We assumed a bin size of 
$\Delta t=10$ ks (so that $T_{\rm min}$ in equation (\ref{eq:sigmabn}) is equal to 20 ksec). This is typical 
for relatively faint sources, where a large bin size is required to increase the 
S/N ratio, hence decreasing the contribution of the experimental Poisson 
noise in the observed variations. As a result, our original lightcurves had 
7800 points in them. We randomly chose a lightcurve segment with a length 
equal to $T_{\rm max}$, and then, we chose these points from this segment 
which reproduce the actual observing pattern of the 1Ms XMM-CDFS observations. 

Figure \ref{fig:lc_sparsyXMM} shows an example of the simulated lightcurve in the case of $\beta=1.5$ PSD. 
The blue continuous line indicates the \textit{full} lightcurve. Black crosses with error bars 
indicate the data points in the lightcurve segment of size $T_{\rm max}$ that we 
randomly chose (hereafter "continuously sampled lightcurve''), while the red points 
indicate the actual XMM-CDFS observations, that were performed within this interval 
(hereafter "sparsely sampled light curve'').

\begin{figure}[]
\begin{center}
\includegraphics[height=6.5cm,bb=100 371 600 760]{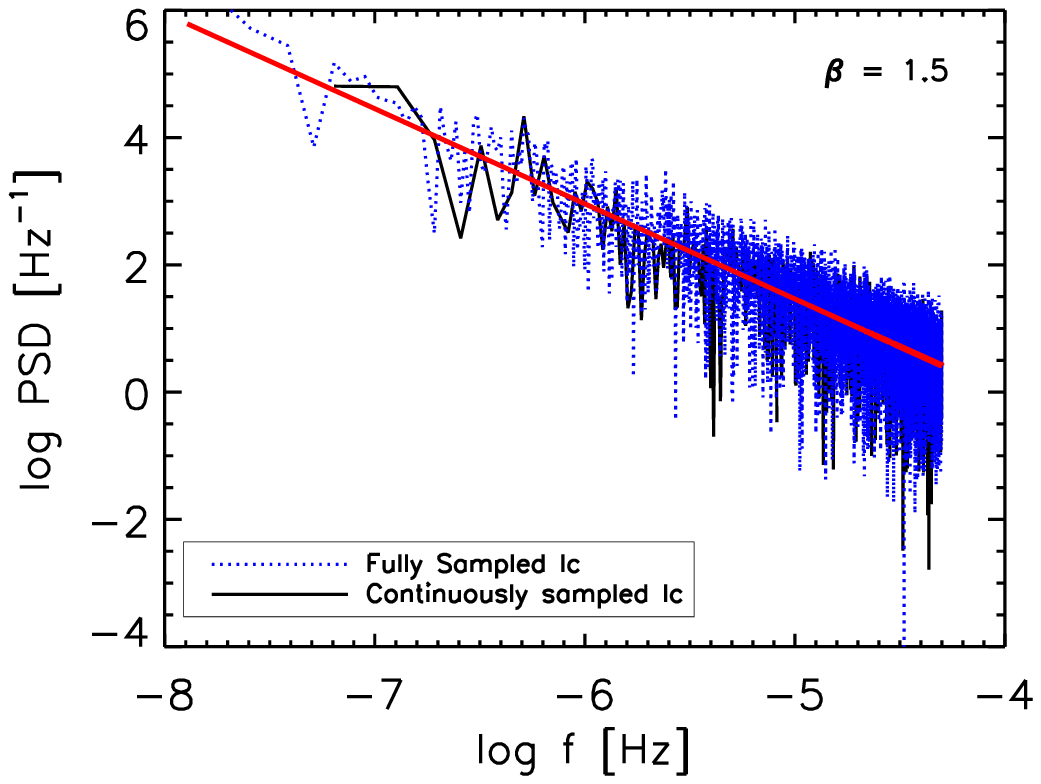}
\caption{Periodogram derived from the lightcurves shown in Figure \ref{fig:lc_sparsyXMM}. The blue dotted line corresponds to the periodogram of the full lightcurve (blue line in Figure \ref{fig:lc_sparsyXMM}), with the red line showing the input PSD. The black line corresponds to the periodogram of the continuously sampled lightcurve (black points in Figure \ref{fig:lc_sparsyXMM}).}
\label{fig:psd}
\end{center}
\end{figure}

The blue dotted line in Figure \ref{fig:psd} indicates the periodogram 
of the full lightcurve indicated with the blue points in Figure \ref{fig:lc_sparsyXMM}. 
The red solid line in the same figure indicates the input PSD, with a slope of $\beta=1.5$. 
The agreement between the periodogram of the full simulated lightcurve and the 
"input'' PSD is very good. The black solid line indicates the periodogram of the "continuous'' 
lightcurve segment (which is indicated with the black points in Figure \ref{fig:lc_sparsyXMM}). 
This is not significantly different than the intrinsic PSD either. This is probably due to 
the fact that red-noise aliasing effects are not extreme in the case of PSDs with slopes less than 2. 

For each one of the 5000 simulated lightcurves (like the one shown in Figure 
\ref{fig:lc_sparsyXMM}) we computed  the sample mean of the randomly chosen 
"continuously'' and "sparsely'' sampled lightcurves ($\overline{x}_{\rm cont}$ 
and $\overline{x}_{\rm sparse}$, respectively) and their normalized excess variance 
using Eq. \ref{eq:exvar2} (\scon\ and \sspa, respectively). We also 
computed the ML variance estimator as proposed by \citet{Almaini} for the 
sparsely sampled lightcurves ($\sigma^2_{\rm ML,sparse}$). Using the 5000 values of  
\scon, \sspa\, and $\sigma^2_{\rm ML,sparse}$, we constructed their sample 
distributions, and we computed their mean values (denoted with brackets, 
i.e. $\langle$ \scon\ $\rangle$, etc), standard deviation, and skewness.

Finally, for each PSD slope $\beta$, we also calculated \sbn\ 
using Eq. (\ref{eq:sigmabn})\footnote{By construction, the intrinsic power-spectrum of the simulated light curves is defined only at a certain 
grid of frequencies, and is not continuous. Therefore, $\sigma^2_{\rm band}$ 
is in reality equal to the sum of the PSD value at each frequency times the 
frequency width, which in our case is equal to $1/(5\times T_{\rm max})$. However, 
we verified that this sum is in effect identical to the integral of the PSD from 
$\nu_{\rm min}$ up to $\nu_{\rm max}$, as defined in Eq. (\ref{eq:sigmabn})}. Since the same equation holds for 
$\sigma^2_{\rm full}$ as well (with $\nu_{\rm min,full}=1/T_{\rm max,full}$, 
and $\nu_{\rm max,full}=1/T_{\rm min}$), one can easily show that: 

\begin{equation}
\label{eq:sigmain_analitical}
\sigma^2_{\rm band,norm}=\sigma^2_{\rm full,norm} \times \frac{(T_{\rm max}/T_{\rm min})^{\beta -1}-1}{(T_{\rm max,full}/T_{\rm min})^{\beta-1}-1},
\end{equation}

in the case of PSDs with $\beta>1$, and: 

\begin{equation}
\label{eq:sigmain_analitical2}
\sigma^2_{\rm band,norm}=\sigma^2_{\rm full,norm} \times \frac{\ln(T_{\rm min}/T_{\rm max})}{\ln(T_{\rm min}/T_{\rm max,full})},
\end{equation} 

in the case of PSDs with $\beta=1$. The \sbn\ values for the different PSD 
slopes are listed in Table 2. Note that, although 
$\sigma^2_{\rm full,norm}$, $T_{\rm max}$ and $T_{\rm min}$ are the same for all 
the simulations we performed, \sbn\ 
decreases significantly with increasing $\beta$, as more power is 
stored in the low frequency components for steeper PSDs.

\begin{table}
  \centering
    \caption{Input parameters of the simulated AGN lightcurves}
      \label{tab:input_param}
\begin{tabular}{l l}
  \hline
  \hline
  Power-law PSD index ($\beta$) & 1,1.5,2,2.5,3 \\
  Number of simulations ($N$)  & 5000 \\
  $\mu_{\rm full}$ & 0.099 cnt s$^{-1}$\\
  Time resolution ($\Delta t$)  & 10 ks \\
  $\sigma^2_{\rm full,norm}$ & $0.042$ \\
  Background level & 0.06 cnt s$^{-1}$ \\
\hline
\end{tabular}
\end{table}

\section{The distribution of the normalized excess variance, in the case of red noise PSDs}
\label{sec:4}

\begin{figure}
  \includegraphics[height=6.5cm]{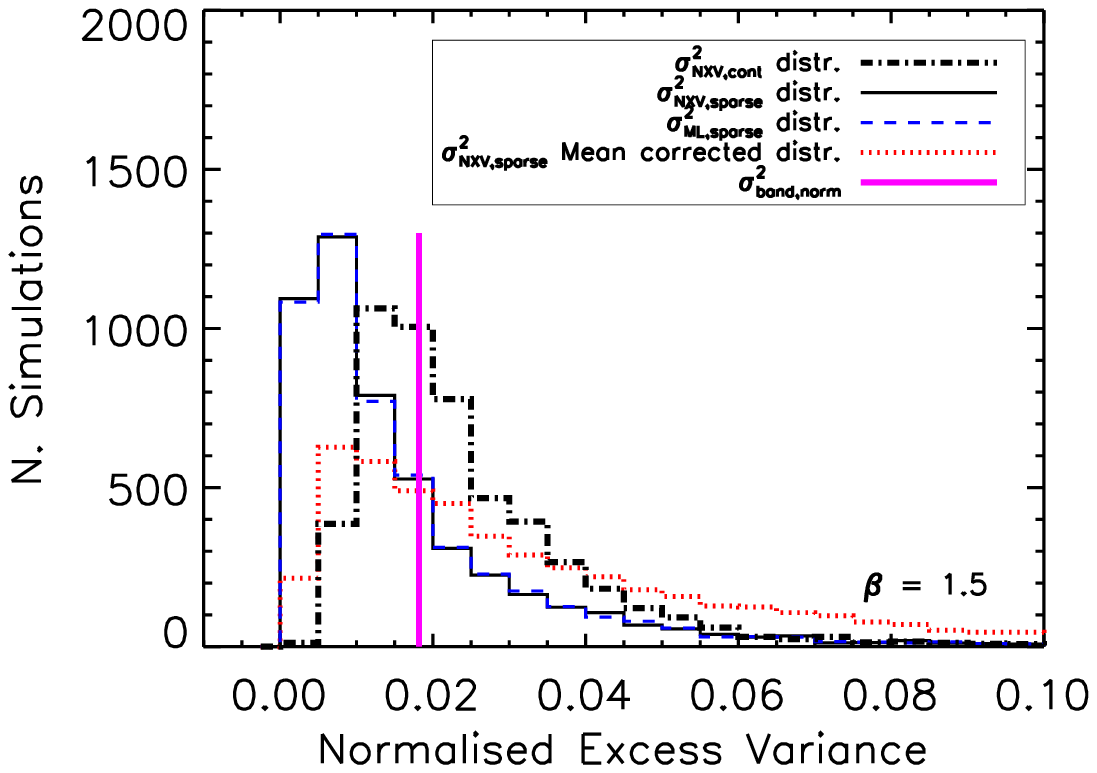} 
\caption{The \sspa, \scon, and  $\sigma^2_{\rm ML,sparse}$ distributions (black solid, black dot-dashed lines, and dashed blue lines, respectively), compared to \sbn\ (vertical red solid line), for the $\beta=1.5$ PSD. The dotted red line indicates the \sspa\ distribution, when \sNXV\ is computed using the intrinsic mean count rate $\mu_{\rm full}$ (see discussion in the text).}
\label{fig:histoexv}
\end{figure}

\begin{table*}[t]
\caption{Statistical properties (mean/standard deviation/skewness), formal error and bias of the variance distributions}
\label{tab:tablemaxlik}
\centering
\begin{tabular}{l l l l l l l l c c}
\hline\vspace{-0.2cm}\\
$\beta$ &$\sigma^2_{\rm band,norm}$& $\sigma^2_{\rm NXV,cont}$ & $\Delta \sigma^{2}_{\rm NXV}$ & $\sigma^2_{\rm NXV,sparse}$ & $\Delta \sigma^{2}_{\rm NXV}$ & $\sigma^2_{\rm ML, sparse}$ & $\sigma^2_{\rm NXV,sparse}$ & $b_{\rm cont}$ & $b_{\rm sparse}$  \\
 &   &   & Eq. 2 &  & Eq. 2 & & Mean Corr. &  & \\
\hline

1 & 0.034 & 0.037/0.01/1.35 & 0.001 &  0.029/0.017/3.35 & 0.006  & 0.029/0.017/3.29 & 0.042/0.026/2.44 & 0.92 & 1.19  \\

1.5 & 0.018 & 0.025/0.016/2.55  & 0.0008 &   0.019/0.025/5.24  &  0.004  & 0.019/0.025/5.16 &  0.043/0.044/2.57 & 0.74 & 1.00 \\

2 & 0.0084 & 0.016/0.016/3.42 & 0.0004 &  0.014/0.023/5.52  &  0.003 &  0.014/0.023/5.48 &  0.042/0.048/2.58  & 0.49 & 0.57 \\

2.5 & 0.0037 & 0.012/0.015/3.81  & 0.0003 &  0.012/0.022/6.03 &  0.003 & 0.012/0.022/6.09 &  0.041/0.048/2.42 & 0.31 & 0.30  \\

3 & 0.0017 & 0.010/0.013/4.11 &  0.0003 & 0.012/0.023/8.12  & 0.003 & 0.012/0.023/8.20 &  0.042/0.051/2.59  & 0.17 & 0.14 \\

\hline

\end{tabular}
\end{table*}

The black dot-dashed, black solid and blue dashed lines in Figure \ref{fig:histoexv} indicate the 
\scon, \sspa, and $\sigma^2_{\rm ML,sparse}$ distributions, using the results from the set of the 
5000 simulations for the case of the $\beta=1.5$ PSD.  
Although the errors on each lightcurve point are not identical (see \S\ref{sec:2}), 
the $\sigma^2_{\rm ML,sparse}$ distribution does not  differ significantly 
from the distribution of the \sspa\ values. 

All distributions are broad and asymmetric. They are skewed, and show long tails 
towards values larger than \sbn\ (which is indicated by the vertical solid line in the 
same plot). This is mainly due to the fact that the excess variance follows a 
$\chi^2_N$ distribution (see Vaughan et al. 2003). At the same time, a large number of 
normalized variance values are quite smaller than \sbn. This is  particularly true 
with the \sspa\ and $\sigma^2_{\rm ML,sparse}$ distributions, because in many 
cases the data points in the sparsely sampled light curves are ``clustered'' close 
to their sample mean, hence resulting in sample variances smaller than the intrinsic value.
\begin{figure}   
\includegraphics[height=13cm,bb=0 0 500 850]{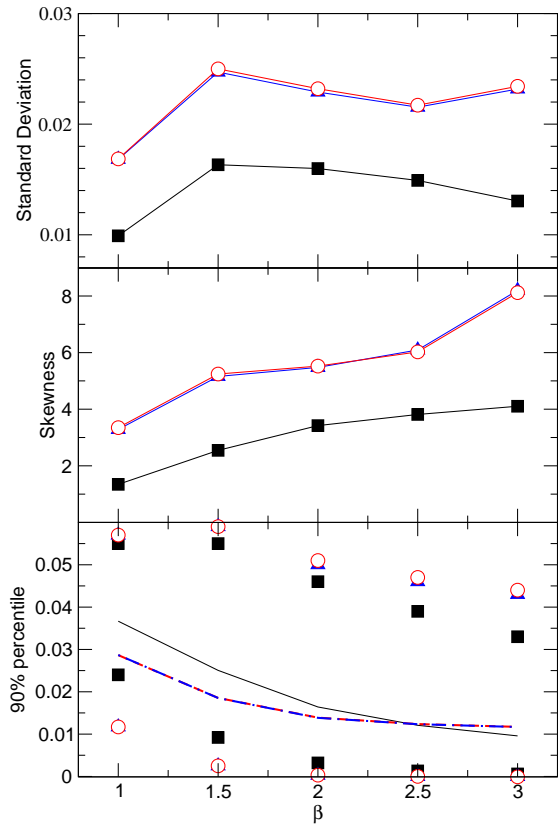} 
\caption{Standard deviation, skewness and 90\% percentiles of the
\scon, \sspa\ and $\sigma^2_{\rm ML,sparse}$ distributions (filled black squares, 
open red circles, and filled blue triangles, respectively) plotted as function of 
$\beta$. In the lower panel the black continuous, red dashed and blue dot-dashed lines represent the 
sample means as reported in Table \ref{tab:tablemaxlik}.}
\label{fig4}
\end{figure}

The mean, standard deviation and skewness of the normalised excess variance distributions 
for all the PSDs we considered are listed in Table \ref{tab:tablemaxlik} (the three numbers 
separated by a slash indicate the mean, standard deviation and skewness 
of each distribution listed on the first line of this table). 

On average, \scon\ measures rather well \sbn\ in the case of the $\beta=1$ PSD. 
However, as $\beta$ increases, $\langle$\scon$\rangle$ becomes larger than \sbn. 
Therefore, the normalized excess variance is 
a biased estimator of \sbn, even in the case of continuously sampled lightcurves. 
The situation is more complicated in the case of the normalized excess 
variance estimates when there are missing points. Although these estimates are also biased, 
we find that $\langle\sigma^2_{\rm NXV/ML,sparse}\rangle$ is {\it smaller} than \sbn\ in the 
case of a PSD with $\beta=1$, $\langle\sigma^2_{\rm NXV/ML,sparse}\rangle \approx \sigma^2_{\rm band,norm}$ 
in the case of the $\beta=1.5$ PSD, and then 
it is larger than \sbn\ for steeper PSDs.  Therefore the bias of the normalized 
excess variance must depend on both the PSD slope and  the sampling pattern of the lightcurve. 

The top panel in Figure \ref{fig4} shows the standard deviation of the 
\scon, \sspa\ and $\sigma^2_{\rm ML,sparse}$ distributions (filled black squares, 
open red circles, and filled blue triangles, respectively) plotted as function of 
$\beta$. The middle and bottom panels in the same figure show the skewness 
and the 90\% quantiles of the distributions, plotted as a function of $\beta$ 
(point markers are as in the top panel of the figure). The solid black line, 
the red dashed line and the dash-dotted blue line in the bottom panel indicate 
the sample mean of the \scon, \sspa\ and $\sigma^2_{\rm ML,sparse}$ distributions, respectively. 

The first thing to notice is that it is very hard to distinguish between the (blue) 
filled triangles and (red) open circles in all panels, as well as the (red) dashed 
and (blue) dot-dashed line in the bottom panel. This result indicates that the 
statistical properties of the \sspa\ and \sML\ distributions are identical in all cases. 
Both methods to compute the normalised excess result in estimates with 
identical statistical properties. For this reason, we will not explicitly consider the 
properties of \sML\ in the following discussion.

The top panel in Figure \ref{fig4} shows that the standard deviation of all distributions 
increases from $\beta=1$ to $\beta=1.5$, 
and then remains roughly constant. However from Table 2 it can be seen that 
the ratio of the standard deviation over the sample mean (i.e. the relative width of the distribution) 
increases with PSD slope. The 
distributions also show an increasing asymmetry as $\beta$ increases. 
The skewness in all cases is positive, and increases with PSD slope. 
The bottom plot in Figure \ref{fig4} conveys similar information. The 90\% quantiles 
are asymmetric with respective to the mean of the distribution, 
and this asymmetry is more pronounced for the distributions 
in the case of the sparsely sampled data. 

Figure \ref{fig4} also shows clearly the effect of uneven sampling to the statistical 
properties of the normalized excess variance. The standard deviation of the \sspa\ 
distributions is larger than the standard deviation of the \scon\ distributions by a factor of 
$\sim 1.8$, at all $\beta$'s. The same holds for skewness: the \sspa\ distributions are significantly more asymetric than the \scon\ distributions for all PSD slopes. 

We also used the simulated lightcurves and Eq.\ref{eq:exvar_err} to calculate the mean 
``error'' of \scon\ and \sspa, for all PSD slope values. We found that, apart from the fact that this error cannot account for the 
asymmetry of the excess variance distributions, it is always smaller than their standard deviation, i.e. the formal error tends 
to underestimate the true scatter of the excess variance estimates. This was also shown by Vaughan et al. (2003) in the 
case of evenly sampled lightcurves. We found that it is also true in the case of the sparsely sample light curves. 
This is not surprising, given the fact that the standard deviation of \sspa, in always larger than the standard deviation of \scon. 

We also examined the statistical properties of \sspa\ distributions, if 
we use the intrinsic mean, $\mu$, of the lightcurves instead of the sample 
mean, $\bar{x}$, when we estimate the normalized excess variance using 
Eq. (\ref{eq:exvar2}).  For this reason, we fixed the average count rate 
$\overline{x}$ in Eq. (\ref{eq:exvar2}) to its intrinsic value, $\mu_{\rm in}=0.1$ 
cnt s$^{-1}$, and we recalculated \sspa\ for the 5000 synthetic lightcurves 
with sparse sampling, for all $\beta$'s. The resulting distribution is 
shown in Figure \ref{fig:histoexv} by the dotted line in the case of $\beta=1.5$.  
The mean, standard deviation and the skewness of the distributions for all 
$\beta$'s are listed in Table 2 under the column $\sigma^2_{\rm NXV}-$ "Mean Corrected''.  
The mean-corrected distribution yields almost always the same mean, 
for all $\beta$'s, which is equal to the input value of the total lightcurve, 
$\sigma^2_{\rm full,norm}$, and not just \sbn\ (Table \ref{tab:input_param}). 
This result is in agreement with the fact that the sample variance is 
an unbiased estimator of the {\it intrinsic} variance of a time series, irrespective 
of the length of the lightcurve, if one knows the true mean of the time series 
in advance. Even in the case of the sparsely sampled lightcurves, since the mean 
is fixed to the intrinsic value, the sparsely sampled points randomly probe the full 
scale of fluctuations around the intrinsic mean $\mu_{\rm in}$, thus yielding, 
on average, the correct variance. 
However, the intrinsic mean of a lightcurve is hardly known in practice, so 
we continue below with the study of the statistical properties of \scon\ and \sspa. 

Finally we considered the aliasing effect discussed in Section \ref{sec:2}, and estimated how it affects our results. To this end we repeated the simulations creating light curves with a bin size of 2 ks, and then estimated the mean of each 5 points, so that we end up with a light curve with a bin size of 10 ksec as before. In this way, we mimic the binning that is performed in real data, where the light curves are not \textit{sampled} but \textit{binned} over intervals of size $\Delta t$. 
We found that the mean (as well as the other statistical moments of the distributions) are almost identical to the ones reported in Table 2, in all cases when $\beta\geq 1.5$. Only for a PSDs with $\beta=1$, we find that $\sigma^2_{NXV}$ is mildly affected by aliasing, which increases the measured variance by $\sim 5\%$ for sparse sampling, and $\sim 10\%$ in all other cases (the bias discussed in next section is reduced accordingly by the same amount). This result indicates that, in this case of a very flat PSD, despite the binning which suppresses the variability at time scales smaller than 10 ksec, there is still some variability power which ``appears'' in the sampled frequency band. However this difference is not significant and does not affect any of our conclusions; furthermore  such effect depends on the PSD slope at the \textit{high} frequency threshold which, for typical X-ray observations, lies on the steep part of the PSD where its importance is negligible.

One of the main results presented in this section show that \sNXV, both for continuously and 
sparsely sampled lightcurves, are biased estimates of \sbn. In the following section 
we present a method to estimate the bias of the normalized excess variances measurements.  

\begin{figure}
   \includegraphics[width=7cm, bb=40 315 508 825]{bias.eps} 
        \caption{Bias factor as a function of $\beta$ (from Table \ref{tab:tablemaxlik}),
        compared with the predictions based on Eq. (\ref{bias_pred_cont}). Black solid squares
        and open red circles represent the bias of \scon\ and \sspa\ respectively, while open green and 
        solid blue diamonds indicate the behavior in the case of \textit{uniform} and \textit{progressive}
        sampling patterns discussed in \S \ref{sec:5.1}.
        The solid line shows the theoretical bias prediction as described in Section 5.3, and the dashed line indicates the same
        relation multiplied by 1.3 for PSDs with slopes flatter than 2 (see text for details).}
\label{fig:biasbetas}
\end{figure}

\section{The $\sigma^2_{\rm NXV}$ bias}
\label{sec:5}

We define the bias, $b$, of the \scon\ and \sspa\ estimates as follows: 

\begin{equation}
\label{eq:b}
 b=\frac{\sigma_{\rm band, norm}^2} {\langle \sigma^2_{\rm NXV, cont.\mbox{\small or}~sparse} \rangle}        
\end{equation}

In essence, $b$ represents the correction factor that, if known, could be 
multiplied with the individual normalized excess variance estimates so 
that they could be considered as unbiased ({\it on average}) estimates of \sbn.

The bias values of \scon\ and \sspa\ are listed in the last columns 
of Table \ref{tab:tablemaxlik} for all the PSD slopes we considered. 
Figure \ref{fig:biasbetas} shows the bias values plotted as a function of 
$\beta$ in the case of \scon\ (black, filled squares) and \sspa\ (open, red circles). 

The bias certainly depends on the PSD slope. For the continuous sampling case, $b$ is always smaller than 
$1$, and decreases with increasing $\beta$.  This is due to the fact that the variance measured on any given timescale 
range will be affected by leakage effects that, in the case of red-noise PSDs, 
tends to add power coming from nearby, lower frequencies to the observed lightcurve segment. 
This "leakage effect" becomes "stronger" with increasing $\beta$. 
For $\beta\le 2$ (PSD slopes which are 
usually observed in radio-quiet Seyfert nuclei) the bias is of the order of 
$5-50$\% of the intrinsic band normalized variance. However, at higher $\beta$
(as found for radio-loud AGN, or for optical light curves, see \S\ref{sec:intro}), the computed 
normalized excess variances can be up to $\sim 3-7$ times larger than \sbn. 

In the case of sparsely sampled lightcurves, the situation is more complicated. The bias factor 
$b_{\rm sparse}$ is {\it not} always smaller than unity.  In fact, it is larger than unity 
for PSDs flatter than $\beta =1.5$, and then becomes smaller than unity but is larger
than $b_{\rm cont}$ for PSD slopes up to $\beta \sim 2 - 2.3$. We believe that this is due to 
the fact that the additional power due to red-noise leakage is compensated 
by "missing'' power due to the gaps in the lightcurve. At even steeper PSDs, 
$b_{\rm sparse}$ becomes smaller than $b_{\rm cont}$. 

We also point out that, as discussed in the previous section, aliasing effect may 
slightly change the exact value of the bias by $5-10\%$ in the case of a flat PSD, but this depends
on the PSD slope at the high frequency threshold, as opposed to the low frequency limit which
is the one relevant to the leakeage effects discussed above.

Our results thus suggest that $b$ depends on the sampling pattern of the lightcurve as well. We investigated 
this issue, by considering different sampling patterns  as follows. 

\subsection{Bias dependence on the sampling pattern: Uniform and Progressive Sampling}
\label{sec:5.1}

\begin{figure}
   \includegraphics[height=6.8cm,bb=100 371 600 760]{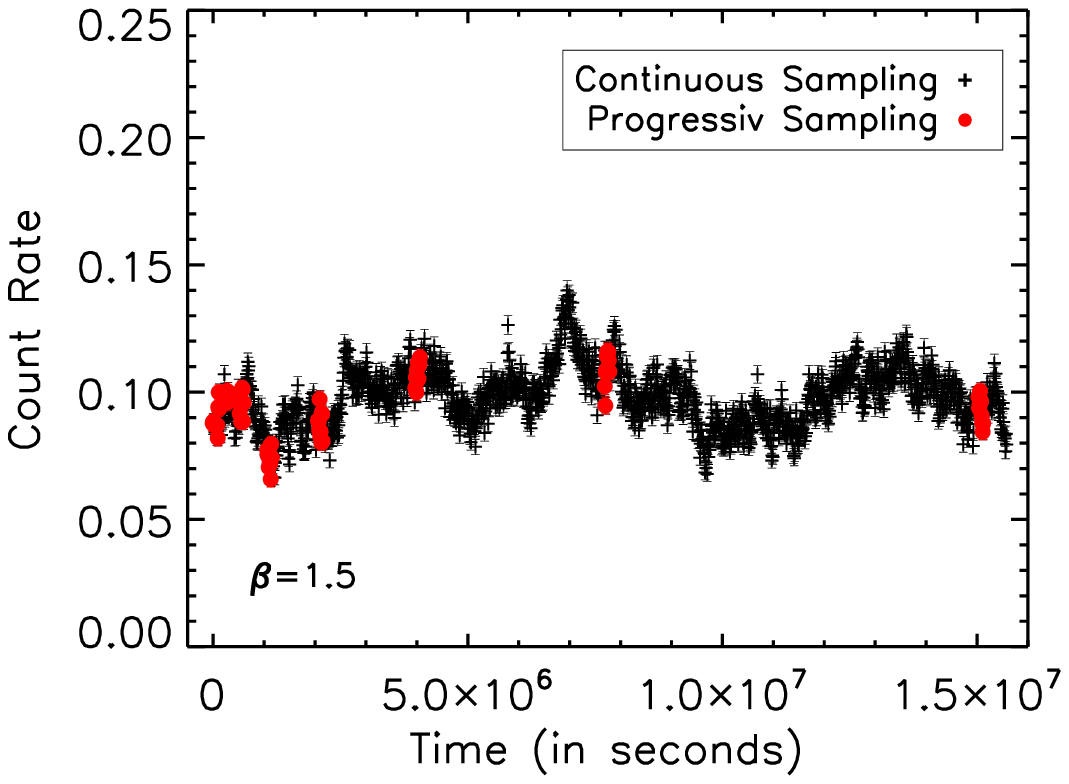} 
   \includegraphics[height=6.8cm,bb=100 371 600 760]{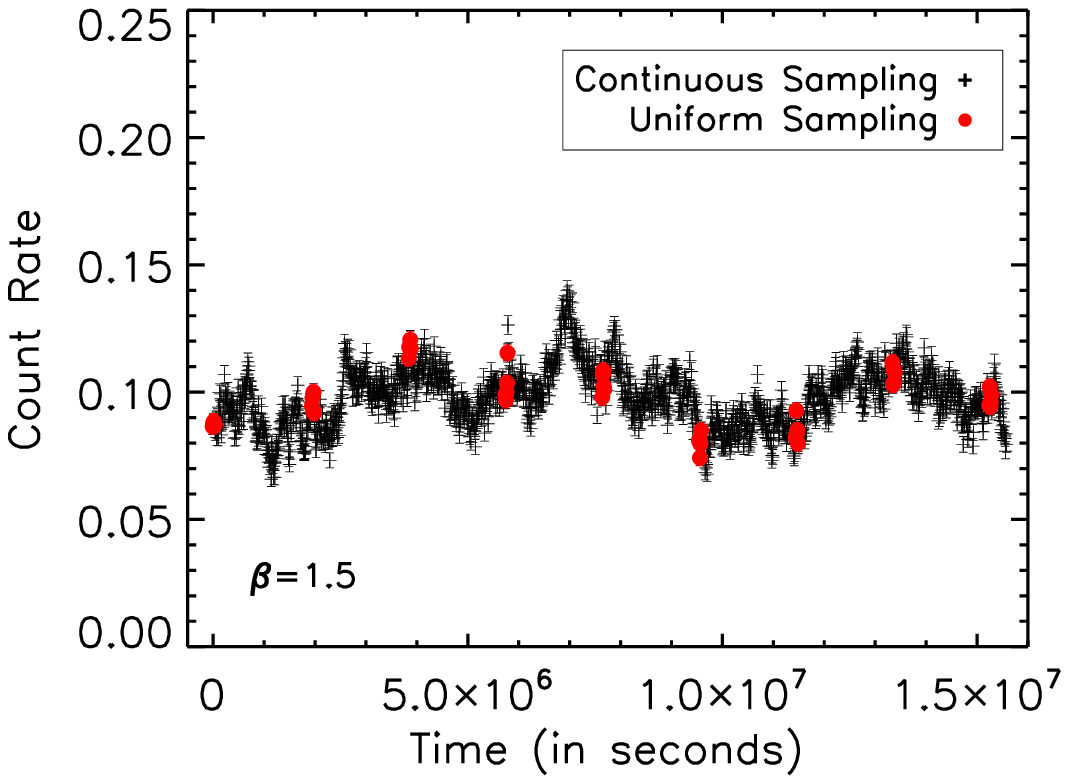} 
      \caption{Simulated continuously sampled AGN lightcurves (black crosses) compared with the progressive (\emph{upper panel}) and uniform (\emph{lower panel}) sampling schemes marked by red circles.}
\label{fig:lc_uni_prog}
\end{figure}

We first examined what would be the bias if we had observed the same object, 
with the same exposure time (i.e. $\sim 450$ ks), over the same period of 
$\sim$ half a year, but instead of performing the observations at the start 
and the end of this time window, they were performed in a more regular way. 
To this end, we used the 5000 lightcurves we had generated with the input parameters 
shown in Table \ref{tab:input_param},  and we then chose from the "continuously 
sampled'' lightcurves the right points the following these sampling patterns:

\begin{enumerate}
\item \textit{Uniform} sampling, consisting in 9 observations of 50 ks each 
          separated by constant temporal gaps of 1900 ks 
          ($\sim 20$ days; Fig. \ref{fig:lc_uni_prog}, \emph{lower panel});
\item \textit{Progressive} sampling, where the observations are separated 
          by increasing lags according to the expression $gap = 2^n \times 10$ ks, 
          with $n=1,2,..,8$ (Fig. \ref{fig:lc_uni_prog}, \emph{upper panel});
\end{enumerate}

\begin{figure}
   \includegraphics[height=6.8cm,bb=100 371 600 760]{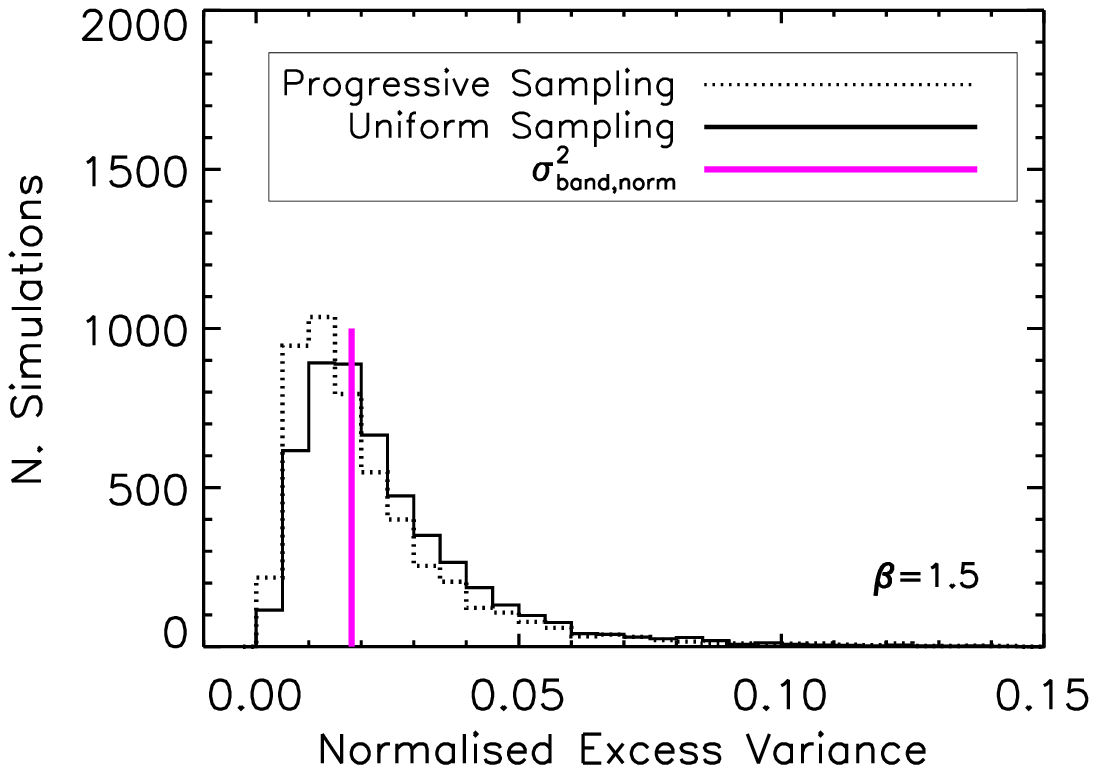} 
      \caption{The distribution of $\sigma^2_{\rm NXV}$ using the N=5000 simulated light curves with uniform (solid line) and progressive (dotted line) sampling. The vertical thick magenta line indicates $\sigma^2_{\rm band,norm}$, like in Fig. 3.}
\label{fig:histo_uni_prog}
\end{figure}

\begin{table}
\begin{center}
\caption{Properties and bias of the $\sigma^2_{\rm NXV}$ distributions for uniform and progressive sampling.}
\label{tab:tableunif}
\begin{tabular}{l l l c c}
\hline\vspace{-0.3cm}\\
$\beta$ & $\sigma^2_{\rm NXV}$ & $\sigma^2_{\rm NXV}$ & $b_{\rm uni}$ &  $b_{\rm pro}$ \\
   & Uniform & Progressive &  & \\
\hline

1 &   0.035/0.016/1.61 &  0.033/0.015/1.38 & 0.97 &    1.02 \\

1.5  & 0.025/0.019/2.27 &  0.022/0.020/5.21 & 0.73 &  0.82 \\

2 &  0.018/0.018/3.42 &  0.016/0.019/5.55 & 0.46 &   0.51 \\

2.5 & 0.013/0.016/3.63  &  0.012/0.018/6.24 & 0.28 &   0.30 \\

3 &  0.011/0.015/4.07  &  0.010/0.016/6.00 & 0.15 &   0.16 \\
\hline
\end{tabular}

\end{center}
\end{table}

Figure \ref{fig:histo_uni_prog} shows the $\sigma^2_{\rm NXV}$  
distribution for the uniform ($\sigma^2_{\rm NXV,uni}$; solid line) and progressive sampling 
scheme ($\sigma^2_{\rm NXV,pro}$; dotted line), derived from the 5000 simulations in the case 
of a PSD with $\beta=1.5$. The vertical solid line indicates \sbn\ 
 for $\beta=1.5$.

The mean, standard deviation and skewness of the distributions, as well 
as the bias values, are listed in Table \ref{tab:tableunif}. The standard 
deviations are similar for both sampling schemes. The skewness 
parameters indicate that the distributions are asymmetric. Skewness is larger in the case of progressive sampling. 
The mean, standard deviation and skewness of the $\sigma^2_{\rm NXV,uni}$ 
distributions are very similar to the same parameters of the \scon\ distributions. 
This is an interesting result, and shows that, even if we do not sample the 
time series continuously, as long as the sampling is uniform, the statistical 
properties of the derived normalized excess variance will be very similar to 
the properties of the normalized excess variance computed from a continuously 
sampled light curve (of similar length). 
Open (green) diamonds and filled (blue) diamonds in Figure \ref{fig:biasbetas} 
indicate the bias, $b$, of the $\sigma^2_{\rm NXV,uni}$ and of 
$\sigma^2_{\rm NXV,pro}$ distributions. The fact that it is difficult 
to discriminate between the filled squares and the open diamonds in 
this plot indicates clearly that the bias factors are identical in the case 
of the continuously sampled and the unevenly sampled lightcurves, when data 
are sampled uniformly. 

This result shows that in general the sampling pattern is 
as important as, if not more important, than the total number of data points. 
The different patterns analyzed here and in Sec.\ref{sec:4} have approximately
the same total number of points (e.g. total exposure time) but different number
of effectively independent (clusters of) points: the sparse pattern has only 
two independent clusters, the progressive one  5-6 independent clusters,
while the uniform one samples the maximum number (9) of
independent clusters and thus samples better the largest-amplitude/
longest-timescale variability trends (which dominate
contributions to $\sigma^2_{NXS}$). Clearly in some instances, if a more detailed 
analysis than just the measure of the total variance, is required
(such as the detailed study of the PSD), sampling more timescales
as done by the progressive pattern may be preferable.

Finally, also in the analysis of the different sampling patterns, we verified that using the maximum-likelihood approach
yields the same results as using the excess variance, as already found for the sparse
sampling case discussed in Sec. \ref{sec:4}.

\subsection{Bias Dependence on the source flux}
\label{sec:5.2}

\begin{table*}[t]
\begin{center} 
\caption{Statistical properties (mean/standard deviation/skewness) and bias of \sNXV\ as a function of S/N ratio for $\beta=1.5$}.
\label{tab:snratio}

\begin{tabular}{c c c c c c c}
\hline\vspace{-0.3cm}\\
Count Rate & $\frac{S}{N}$ & Flux & $\sigma^2_{\rm NXV}$ & $\sigma^2_{\rm NXV}$ & $b_{\rm con}$ & $b_{\rm spa}$\\
cnt s$^{-1}$ & & \begin{small}
(erg s$^{-1}$cm$^{-2}$)
\end{small} & Continuous & Sparse \\
\hline
0.1  & 25 & 6.25  $\times 10^{-13}$ & 0.025/0.016/2.42 & 0.018/0.023/4.94 & 0.75 & 1.0 \\
0.05  & 22.6 & 3.12  $\times 10^{-13}$ & 0.025/0.016/2.53  & 0.018/0.024/5.36 & 0.75 & 1.0 \\
0.01  & 6.3 & 6.25 $\times 10^{-14}$ &  0.025/0.016/2.52 & 0.017/0.025/5.85 & 0.75 & 1.1 \\
0.005 & 3.4 & 3.12 $\times 10^{-14}$ & 0.025/0.017/2.68 & 0.017/0.033/4.03 & 0.75 & 1.1 \\
0.002 & 1.4 & 1.25 $\times 10^{-14}$ & 0.025/0.024/1.16 & 0.015/0.11/0.94 & 0.75 & 1.2 \\
0.001 & 0.8 & 6.25 $\times 10^{-15}$ & 0.025/0.070/ 0.34 & 0.014/0.43/0.96 & 0.75 & 1.3 \\

\hline
\end{tabular}
\end{center}
\end{table*}

We also examined whether the bias depends on the source flux (or more precisely, 
on the signal-to-noise ratio of the source) as a result of the increased white noise introduced 
by Poisson fluctuations, as the S/N ratio decreases. To estimate such effects, we simulated lightcurves  
assuming different average count rates, corresponding to fluxes smaller
than the one of the source n.68, as is the case for the bulk of the AGN population 
detected in the CDFS. Conversion factors from counts to fluxes were calculated 
assuming a power law spectrum with $\Gamma=1.4$ and 
$N_{\rm H}=8 \times 10^{19}$ cm$^{-2}$.

In Table \ref{tab:snratio} we list the statistical properties of the \scon\ and 
\sspa\ distributions in the case of a $\beta=1.5$ PSD, using the 5000 simulations 
in the case of the XMM-CDFS observation pattern. Our results show that the statistical 
properties of the \scon\ and \sspa\ distributions do not depend on the S/N of the lightcurve, 
as long as S/N $\ge 3$. Obviously, the uncertainty of the individual normalized 
excess variance measurements cannot be reduced just by increasing the S/N ratio of the 
lightcurve (say by increasing its bin size). 

At smaller S/N ratios, the  standard deviation of both distributions increases. This is an important result, 
specially in the case of continuously sample lightcurves. Fixed confidence intervals for the 
variance estimates (like the ones in Table 1 of Vaughan et al, 2003) should only be used for 
``bright'' sources, where the S/N ratio of the available lightcurves is larger than 3. In the case of 
the sparsely sampled lightcurves, the standard deviation of the distribution becomes so large 
that the use of individual \sspa\ estimates will be meaningless. The bias of $\langle$\sspa$\rangle$ 
increases as well, although we cannot be certain whether this is a real effect, or it is due to 
the fact that the standard deviation of the distribution has increased significantly, and 
as a result more simulations may be necessary in order to establish its mean reliably.
We verified that the same result holds for all the PSDs we considered in the previous sections as well. 

\subsection{A simple prescription to estimate the bias of the sample excess variance}
\label{sec:5.3}

As we mentioned above, in the case of red-noise lightcurves,
power leakage from lower frequencies, outside the sampled range,
will result in \scon\ to be a \textit{biased} estimate of \sbn. 
As our results show, $\langle$\scon$\rangle$ systematically 
\textit{overestimates} the intrinsic \sbn, for all PSDs. It seems then 
reasonable to assume that, the normalized excess variance still 
measures some kind of an intrinsic, normalized variance, and that: 

\begin{equation}
\label{eq:integral_cont}
\langle \sigma^2_{\rm NXV,cont} \rangle =\left[ \int_{1/T^\prime_{\rm max}}^{1/T_{\rm min}} PSD(\nu) d\nu \right] /\mu^2
\end{equation}

where $T^\prime_{\rm max} > T_{\rm max}$, to account for the "red-noise leakage" 
effects. Using the above equation, Eq. (\ref{eq:sigmabn}) for $\sigma^2_{\rm band,norm}$, and 
Eq. (\ref{eq:sigmain_analitical2}) for the definition of bias, 
we can show that the bias for continuously sampled lightcurves will be equal to: 

\begin{equation}
\label{bias_pred_cont}
b_{\rm cont} = \frac{T_{\rm min}^{(\beta-1)}-T_{\rm max}^{(\beta-1)}}{T_{\rm min}^{(\beta-1)}-{T^\prime_{\rm max}}^{(\beta-1)}}
\approx \left(\frac{T_{\rm max}}{T^\prime_{\rm max}}\right)^{\beta-1},
\end{equation}

where the final expression assumes that $T_{\rm max},T^\prime_{\rm max}\gg T_{\rm min}$. 
The above equation can also be written as: 

\begin{equation}
\label{bias_pred_cont2}
{\rm log}  \left( \frac{T_{\rm max}}{T^\prime_{\rm max}} \right) = \frac{{\rm log} (b_{\rm cont})}{\beta -1}
\end{equation}

Using the values of $b_{\rm cont}$ listed in Table 2 (for $\beta>1$) we found 
that $T_{\rm max}/T^\prime_{\rm max}=0.48$. The solid line in Figure \ref{fig:biasbetas} 
shows a plot of the function $b_{\rm cont.}=0.48^{(\beta-1)}$. The agreement between 
this line and the data in the case of continuous, uniform and even progressive 
sampling is reasonably good.  The difference between the predicted bias and 
the one determined from the simulations explained in \S\ref{sec:3} in the case 
of continuous sampling is less than 9\% for $1\le \beta \le 2.5$ and increases to 
15\% for $\beta = 3$ (this is an indication that $T^{\prime}_{\rm max}$ itself may be a function of $\beta$). 

These results suggest that, in the case of continuously sampled light curves, 
with no missing points, the bias to the normalized excess variance can be accounted for 
if we assume that power at all frequencies down to approximately half the lowest sampled frequency 
contributes to the observed variability in the light curve. It appears that, for any light curve length 
(i.e. for any $T_{max}$), $\sigma^2_{NXV,cont}$ measures $\sigma^2_{band,norm}$ {\it plus} an 
amount of power which is equal to the integral of the PSD from $1/T_{max}$ to $\sim 1/(2T_{max})$. 
Thus the overall PSD shape at frequencies below $1/T^\prime_{max}$ does not appear to affect significantly the bias of the 
normalized excess variance. As long as we use the local mean of the light curve in the estimation of the 
normalized excess variance, it is the variability components with frequencies ``just'' below $1/T_{max}$ 
that transfer power to the observed light curves, and not the components with frequencies much lower than $1/T_{max}$.

This result allows us to consider the cases when the PSD is not just a simple power law at all frequencies 
higher than $1/T^\prime_{max}$. For example, if there exists a frequency ``break'' in the PSD, where the slope 
changes from $\beta\sim 1$ (at low frequencies) to $\beta\sim 2$ at high frequencies (just like the AGN X-rays PSDs) 
then, our bias prescription could in principle work in this case as well, provided we can make an assumption about 
the PSD slope at frequencies below the lowest sampled frequency, $1/T_{max}$. If the break time scale is shorter 
than $T_{max}$, and one can assume that the PSD slope at frequencies below $1/T_{max}$, down to $1/T'_{max}$, 
is $\sim -1$, then the bias of the normalized excess variance should be minimal. If on the other hand, the break time 
scale is larger than $T'_{max}$,  then we believe that the adoption of a factor $\sim 0.48$ (which is valid in the case 
of power-law like PSDs with $\beta=2$), should provide a reasonable estimate of the bias of the normalized excess 
variance in this case. In general, as long as the PSD has a power-law shape of slope $\beta$ at frequencies below 
$1/T_{max}$, down to $1/T'_{max}(=2.1T_{max})$, then a rough correction for the bias of the normalized excess 
variance (in the case of continuous sampled lightcurves) could be given by:
\begin{equation}
\label{eq:biascor}
\sigma^2_{\rm NXV,cont} ({\rm bias-corr.})=\sigma^2_{\rm NXV,cont.} \times 0.48^{(\beta-1)}
\end{equation}
where $1\le \beta\le 3$. Strictly speaking, we have demonstrated that this 
is the case when the PSD has a simple, power-law like shape at all frequencies 
down to $1/T'_{max}$, but we believe that this prescription should work reasonably 
well, even if the PSD slope steepens (changes) at frequencies higher than $1/T_{max}$.

In the case of discontinously sampled lightcurves, there will be timescales that 
are not sampled ``properly''. Therefore, it seems reasonable to assume that, 
in this case, the normalized excess variance should be an estimate of the 
contribution to the total normalized variance of only these variability 
components that have been sampled. However, it is not easy to determine 
the longest and shorter time scales that have not been sampled in the 
observed lightcurve, as this depends on the sampling pattern in a complicated way. 
Our results indicate that, even if there is a large percentage of missing points 
(in the case of ``uniform sampling'' we discussed in Section \ref{sec:5.1}, the 
percentage of missing points is almost 97\%), the bias of the normalized 
excess variance should be  almost equal to the bias of \scon, as long as the 
data have been sampled in a quasi-evenly pattern. 

In the case of the most 
extreme sampling pattern (for example when the data have been sampled only at the 
beginning and the end of the observing window), the bias of the excess 
normalized variance is larger than $b_{\rm cont}$ by a factor of $\sim 1.3$ 
in the case when $\beta\sim 1-1.5$. It then becomes almost equal to $b_{\rm cont}$ for PSDs with 
$\beta>2$. The dashed line in Figure \ref{fig:biasbetas} indicates the $b_{\rm cont}$ line multiplied by a factor 
of 1.3 for PSDs with slopes less than 2. The difference of the "predicted" and observed bias factors is less than 10\%. 

We remind the reader that aliasing is not explicitly included in the bias values quoted above,
and is not accounted for in equation \ref{eq:biascor}. This approach is correct for most  PSD slopes
and typical sampled timescales, as the results will differ only if the slope is $\beta \lesssim 1$ at the 
maximum sampled frequency (e.g. cases where the PSD is very flat or we are only sampling
long timescales below the ``break'' frequency). 
In particular for $\beta=1$ the bias will be reduced by $5-10\%$ with respect to the values quoted above,
depending on the exact sampling pattern.

\section{Excess Variance measurements in practice}
\label{sec:6}

In the previous sections we provided 
prescriptions to correct for the bias of \scon\ and \sspa. However, although multiplication of 
the sampled normalized excess variance with the appropriate factors
(see Eq. (\ref{eq:biascor})) can result in an unbiased 
measurement of \sbn, given the large width of the 
distribution functions of these estimates, each individual \sNXV\ measurement 
will still be a highly unreliable estimate of \sbn. This is particularly true for the case of sparsely sampled light curves. 

\begin{figure}
   \includegraphics[height=6.3cm,bb=70 371 600 760]{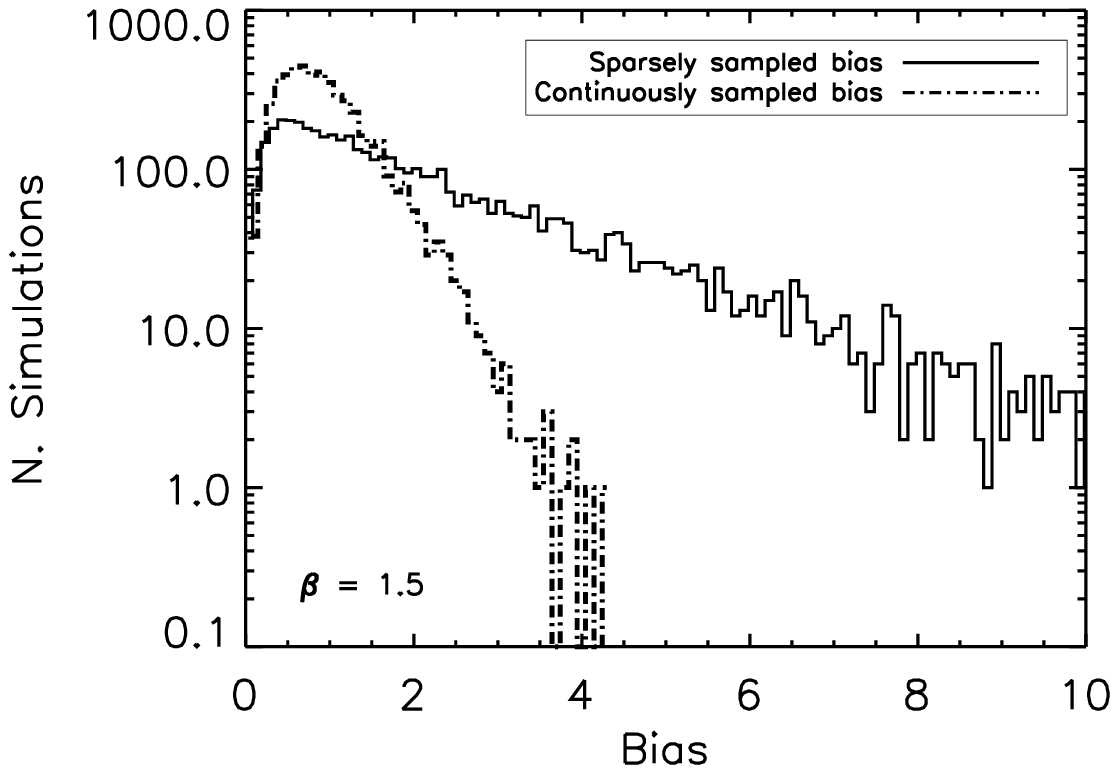} 
        \caption{Bias distribution (of individual measurements) based on a set of 5000 simulated AGN lightcurves such as shown in Fig. \ref{fig:lc_sparsyXMM}, 
        reproducing the continuous and sparse sampling pattern of the {\it XMM-Newton} observation of the CDFS. }
\label{fig:histobias}
\end{figure}

To demonstrate this issue, the solid and dot-dashed lines in Figure \ref{fig:histobias} indicate the bias distribution 
of individual measurements (i.e. the bias as defined in Eq. \ref{eq:b} but 
without using an average value for $\sigma^2_{\rm NXV}$) from the continuously and sparsely sampled 
5000 simulated lightcurves for the case of an intrinsic power-law 
PSD with $\beta=1.5$. These distributions convey the same information as the \scon\ and \sspa\ distributions, which are plotted in Figure 3, and the same comments about their width, and asymmetry, hold for them as well. But perhaps by plotting the distribution of the "bias" factors, it becomes clearer that extreme care should by employed when inferring variability parameters from single observations of AGN, particularly if the lightcurves are sparsely sampled, in a non-uniform pattern.  Although in the continuous sampling case most of the measurements are within a factor of $\sim 2$ from the intrinsic excess variance,  
in the sparse sampling 40\%, 30\% and 20\% of the measurements have 
a bias larger than 2, 3 and 4, respectively. 
Even if the normalized excess variance distributions were multiplied by the  appropriate bias factors so 
that their mean would be close to the intrinsic \sbn, individual \sspa\ estimates may still be significantly different than the intrinsic value. 

We show below that, in order to derive a more robust estimate of the 
intrinsic source variance, we need to collect repeated observations of the 
same source or to use large samples of sources (assuming they have the 
same variability properties) in order to compute "ensemble'' estimates, 
which have more favourable statistical properties. 

\begin{figure}
   \includegraphics[height=6.5cm,bb=100 371 600 760]{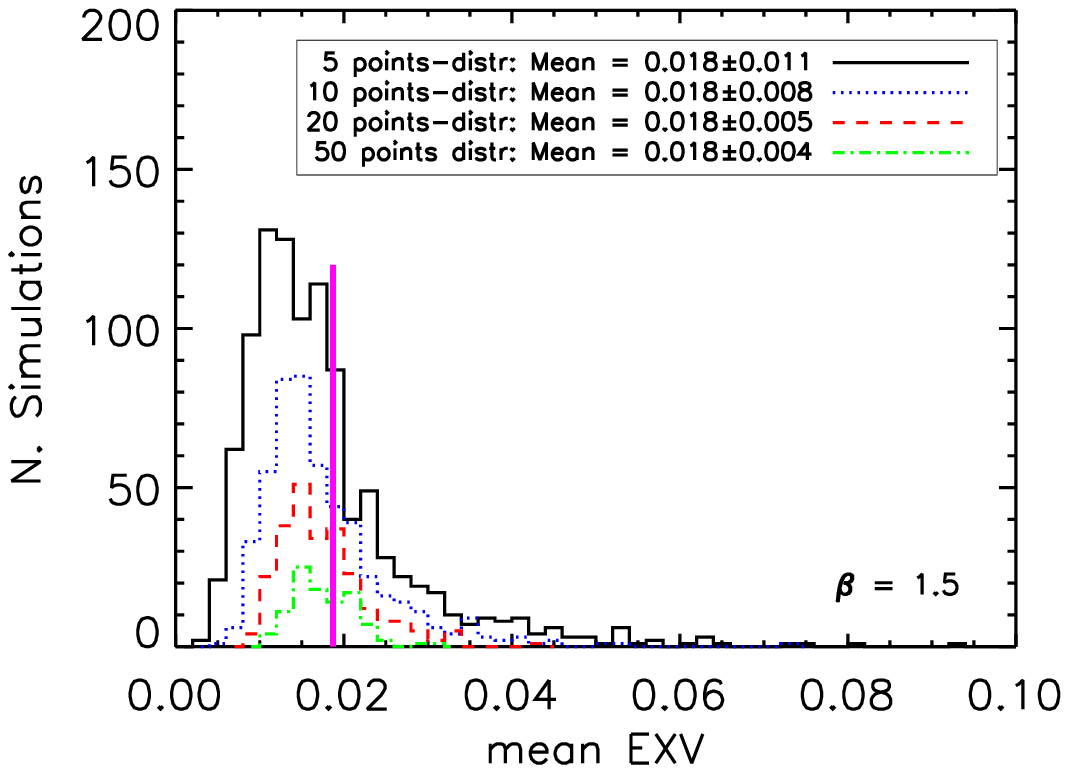} 
   \includegraphics[height=6.5cm,bb=100 371 600 760]{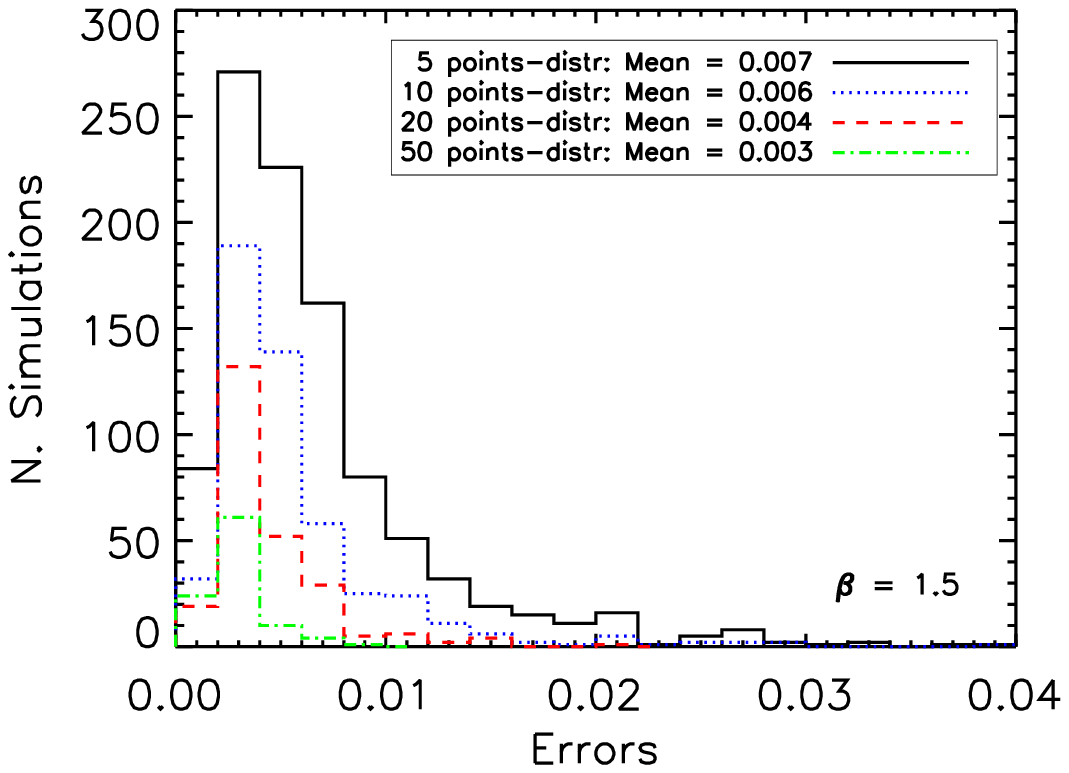} 
      \caption{\textit{Upper panel}: Distribution of \sNXV\ (eq. (\ref{eq:ensemble_nxv})) estimated by binning 5000 simulated excess variance (adopting the XMM sampling pattern, in groups of 5, 10 20 and 50 points (according to the legend). The inset shows the mean values of the binned distributions and their standard deviation, while the vertical thick magenta line shows the intrinsic variance \sbn. The simulations are performed by assuming a count rate of 0.1 cnt s$^{-1}$ and $\beta$ = 1.5. \textit{Lower panel}: Distribution of the errors on $\overline{\sigma^2_{\rm NXV}}$ estimated from Eq. (\ref{eq:ensemble_err}). The inset reports the mean values of the distributions for the different binning.}
\label{fig:binningSN25}
\end{figure}

\begin{figure}
   \includegraphics[height=6.5cm,bb=100 371 600 760]{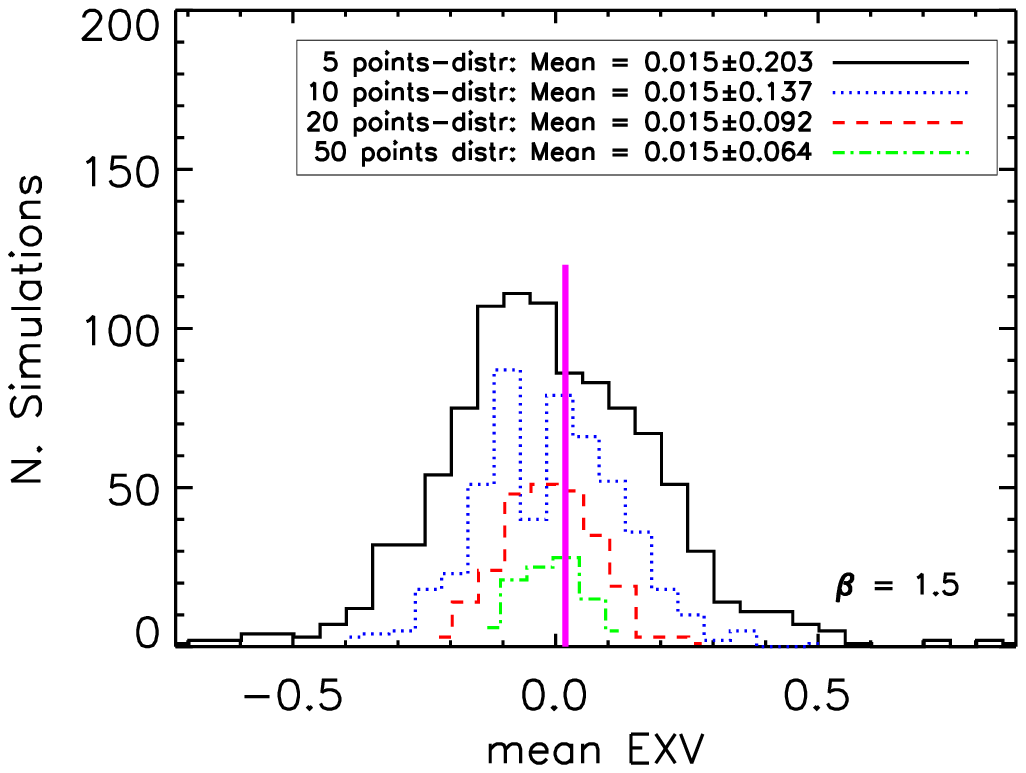} 
   \includegraphics[height=6.5cm,bb=100 371 600 760]{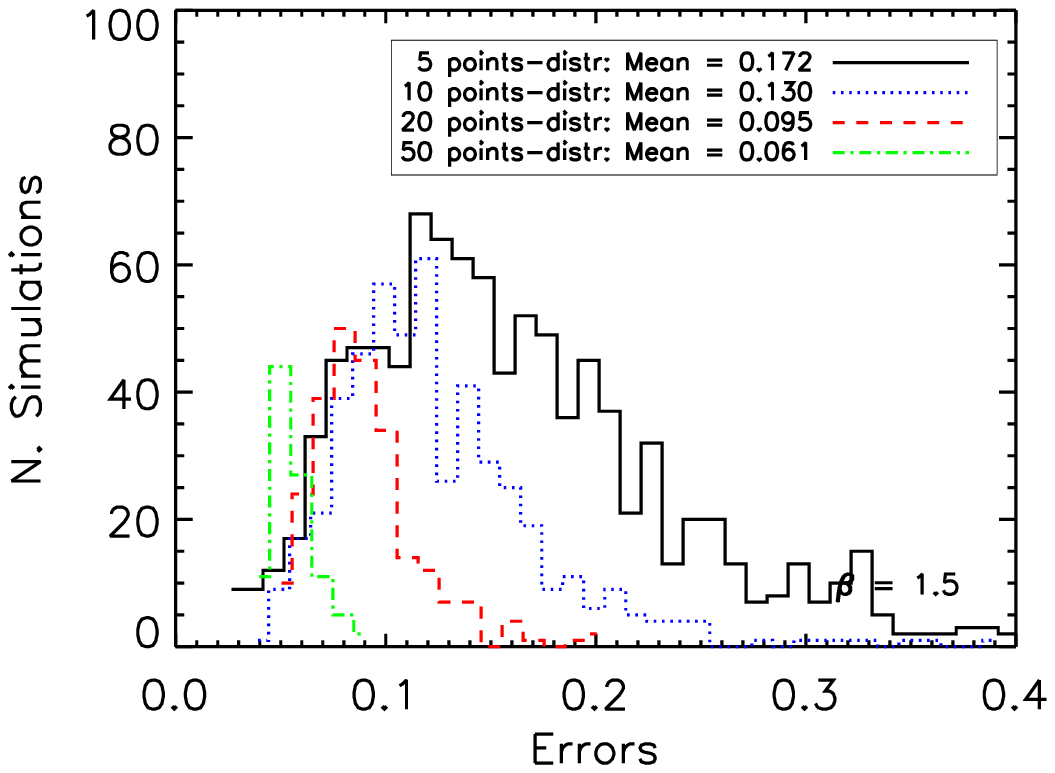} 
      \caption{As Figure \ref{fig:binningSN25} but for 0.001 cnt s$^{-1}$.}
\label{fig:binningSN0.8}
\end{figure}

\subsection{The Ensemble Excess Variance Estimates}
\label{sec:6.1}

We binned the 5000 simulated values of \sspa\  
obtained by using the XMM pattern as described in \S\ref{sec:3} with $\beta=1.5$ 
in groups of $n=5, 10, 20,$ and 50 points. For each bin we estimated the "mean-$\sigma^2_{\rm NXV}$",

\begin{equation}\label{eq:ensemble_nxv}
\overline{\sigma^2_{\rm NXV}}=\sum_{i=1}^n\sigma^2_{\rm NXV, \textit{i}}/n
\end{equation}
and its ``error": 
\begin{equation}\label{eq:ensemble_err}
err(\overline{\sigma^2_{\rm NXV}})=\sqrt{\sum_{i=1}^n [\sigma^2_{\rm NXV,\textit{i}}-\overline{\sigma^2_{\rm NXV}}]/[n(n-1)]}.
\end{equation} 

Note that we dropped the subscript {\it sparse} in the equations above, as our results are applicable to the case of the continuously sampled lightcurves as well. The distributions of the 5, 10, 20 and 50-points binned $\overline{\sigma^2_{\rm NXV}}$ 
are shown in the \textit{upper panel} of Figure \ref{fig:binningSN25} for a count rate of 0.1 
cnt s$^{-1}$ (S/N=25) and $\beta=1.5$. The numbers in the inset window in this panel indicate the mean and standard deviation of each distribution. The mean of these distributions is identical to the mean of the \sspa\ distribution in the case when $\beta=1.5$, but their standard deviation is significantly smaller than the standard deviation of \sspa\, and the distributions are more symmetric. 
In fact, a Kolmogorov-Smirnov test performed on the 5, 10, 20 and 50-points 
mean-$\sigma^2_{\rm NXV}$ distributions indicates that only for the 
5-points grouping we can reject the hypothesis of Gaussian distribution at $>95\%$ level. 

Figure \ref{fig:binningSN25} (\textit{lower panel}) shows the distribution of 
$err(\overline{\sigma^2_{\rm NXV}})$. The mean value of these distributions, which
are quoted in the inset window, are almost identical to  
the standard deviation of the $\overline{\sigma^2_{\rm NXV}}$ distributions for $n\ge 10$. 
We verified that this is the case irrespective on the PSD slope $\beta$.
This shows that the error on 
$\overline{\sigma^2_{\rm NXV}}$, calculated using Eq. (\ref{eq:ensemble_err}), is indeed representative of the true 
scatter of these values around their mean. Therefore, when 
binning the individual $\sigma^2_{\rm NXV}$ estimates in practice, 
we can estimate the intrinsic uncertainty on $\overline{\sigma^2_{\rm NXV}}$ directly 
from the scatter of the individual points around $\overline{\sigma^2_{\rm NXV}}$ in each bin.

Similar results hold if we assume lightcurves with low S/N ratio. In Figure \ref{fig:binningSN0.8} 
we present the distribution of the 5, 10, 20 and 50-points $\overline{\sigma^2_{\rm NXV}}$ 
values, in the case of a lightcurve with a signal-to-noise ratio of 0.8 (i.e. $\sim 30$ times smaller 
than the S/N ratio of the object we considered above). Their mean values are very similar to 
the mean of the respective distributions for brighter sources. Their standard deviation, for 
$n\ge 20$, is at least 5 times smaller than the standard deviation of the distribution of the 
individual \sspa\ estimates (as listed in the bottom row of Table 4).  
Therefore, binning the individual excess variance estimates of faint sources (S/N$\sim 1$) 
by more than $n=20$, allows to retrieve and constrain much better the intrinsic ``signal".
In addition, the 10, 20 and 50-points $\overline{\sigma^2_{\rm NXV}}$ 
distributions are approximately Gaussian, and 
their standard deviation is well approximated by the error of each binned 
estimate (for $n\ge 10$).

But of course, working with faint sources comes at a price. The standard 
deviation of the $n\ge 10$ $\overline{\sigma_{\rm NXV}^2}$ distributions for faint sources is $\sim 16-18$ 
times larger than the standard deviation of the respective distributions for brighter objects. 
Our results indicate that, for a $\sim 3\sigma$ detection of the variability amplitude amplitude in the case of lightcurves 
with S/N ratio $\sim 1$, one will need to bin at least 50 individual normalized excess variance estimates.


\begin{table*}
\centering
\caption{Statistical properties (mean/standard deviation/skewness) and bias of \sNXV\ as a function of S/N ratio for a future mission described in \S \ref{sec:7}}
\label{tab:mcr_exv_bias}
\begin{tabular}{c c c l l l c c}
\hline\vspace{-0.3cm}\\
Mcr\tablenotemark{1} & $\frac{S}{N}$ & Source Flux & $\sigma^2_{\rm NXV}$ & $\sigma^2_{\rm NXV}$ & $b_{\rm con}$ & $b_{\rm spa}$ \\
cnt s$^{-1}$ & &
\begin{small}
(erg s$^{-1}$cm$^{-2}$)
\end{small} & Continuous & Sparse & & \\
\hline
0.1 & 38 & $4 \times 10^{-14}$ & 0.025/0.016/2.43 & 0.024/0.019/2.33 & 0.73 & 0.76 \\
0.01 & 9.3 & $4 \times 10^{-15}$ & 0.025/0.016/2.43 & 0.024/0.021/1.98 & 0.73 & 0.75 \\
0.005 & 7.2 & $2 \times 10^{-15}$ & 0.025/0.017/2.20 & 0.023/0.03/1.54 & 0.73 & 0.79 \\
0.002 & 3.9 & $8 \times 10^{-16}$ & 0.025/0.025/1.18 & 0.020/0.13/1.38 & 0.73 & 0.91 \\
0.001 & 2.7 & $4 \times 10^{-16}$ & 0.025/0.073/0.49 & 0.020/0.49/1.51 & 0.74 & 0.91\\ 
\hline
\end{tabular}
\tablenotetext{1}{\centering\textbf{Mcr}: Mean count rate}
\end{table*}

\section{Constraints on the observing strategy of future X-ray surveys}
\label{sec:7}

Several missions have been proposed over the past few years
to study high redshift AGNs; most of these are designed to have 
larger effective area than current X-ray missions, wider Field-of-View (FOV)
and, depending on the planned orbit, lower background.
For instance the \emph{International X-ray Observatory} \cite[IXO,][]{Barcons2011}
and its evolution \emph{Athena}\footnote{\emph{http://www.mpe.mpg.de/athena/workshop\_mpe\_2011/index.php}},
the \emph{Wide Field X-ray Telescope}  \citep[WFXT,][]{Mur10},
all represent missions capable of performing AGN surveys with higher
speed than \textit{Chandra} or XMM.
The results discussed in the previous sections
allow to explore the capabilities of such future X-ray missions in the
time domain. In particular we examine the expectations for deep, wide-area surveys, 
which will allow to probe the highest redshift and faintest AGN 
populations at the expense of a continuous temporal coverage.

\begin{figure}
   \includegraphics[height=6cm]{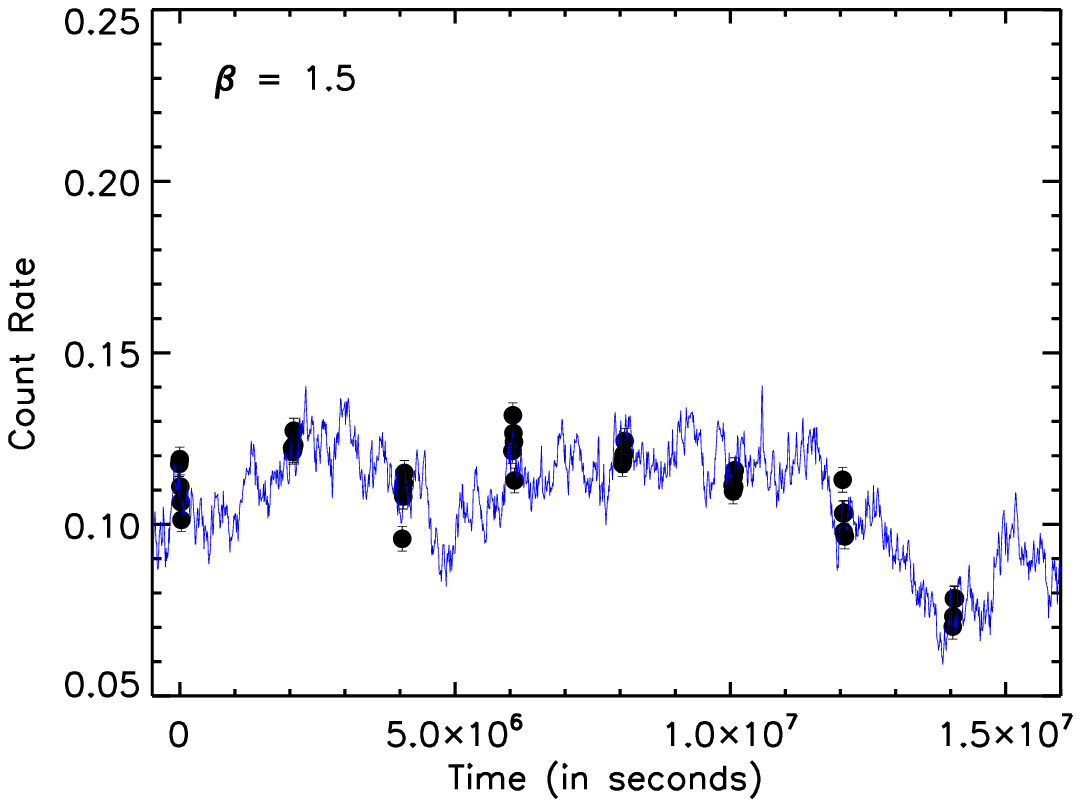} 
      \caption{ Simulated AGN lightcurve, sampled in 50 ks observations spread uniformly on $\sim 6$ months,
      as expected from future large effective area mission such as those described in the text.}
\label{fig:lcWFXT}
\end{figure}

To investigate the capabilities of such missions in measuring AGN variability, we 
present here the performance of a mission with 1 m$^2$ effective area,
1 sq.deg. FOV and the low background allowed by a low earth orbit,
very similar to the WFXT design \citep{Ros11}.
This results in a large number of moderate 
and high redshift AGN \citep[see e.g.][]{Pao11}.
We used a total observing time of $\sim 400$ ks and we evaluated the performance that can 
be expected assuming a uniform sampling scheme very similar to the one presented in \S \ref{sec:5.1}
(although not identical to the due to the WFXT survey constraints).
Figures \ref{fig:lcWFXT} represents an example of a possible observing 
scheme for the survey, where observations of 50 ks each are spread 
evenly over $\sim 6$ months and the corresponding excesses variance and bias distributions,
respectively.

\begin{figure}
   \includegraphics[height=6cm]{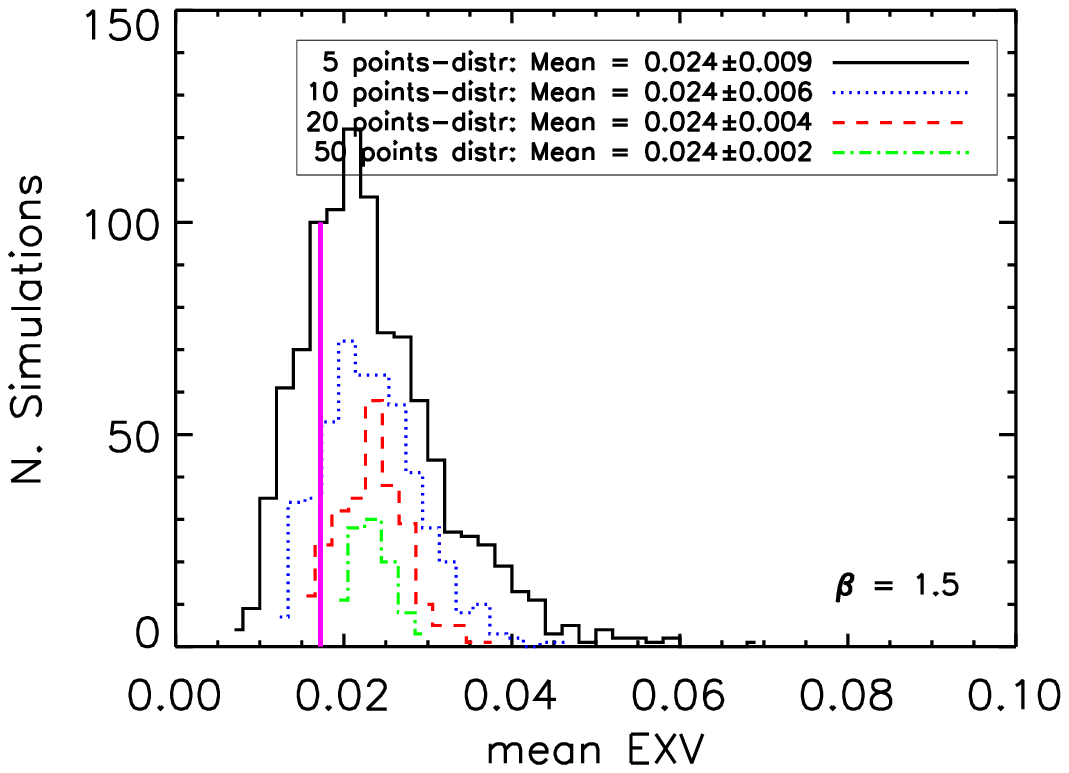}
   \includegraphics[height=6cm]{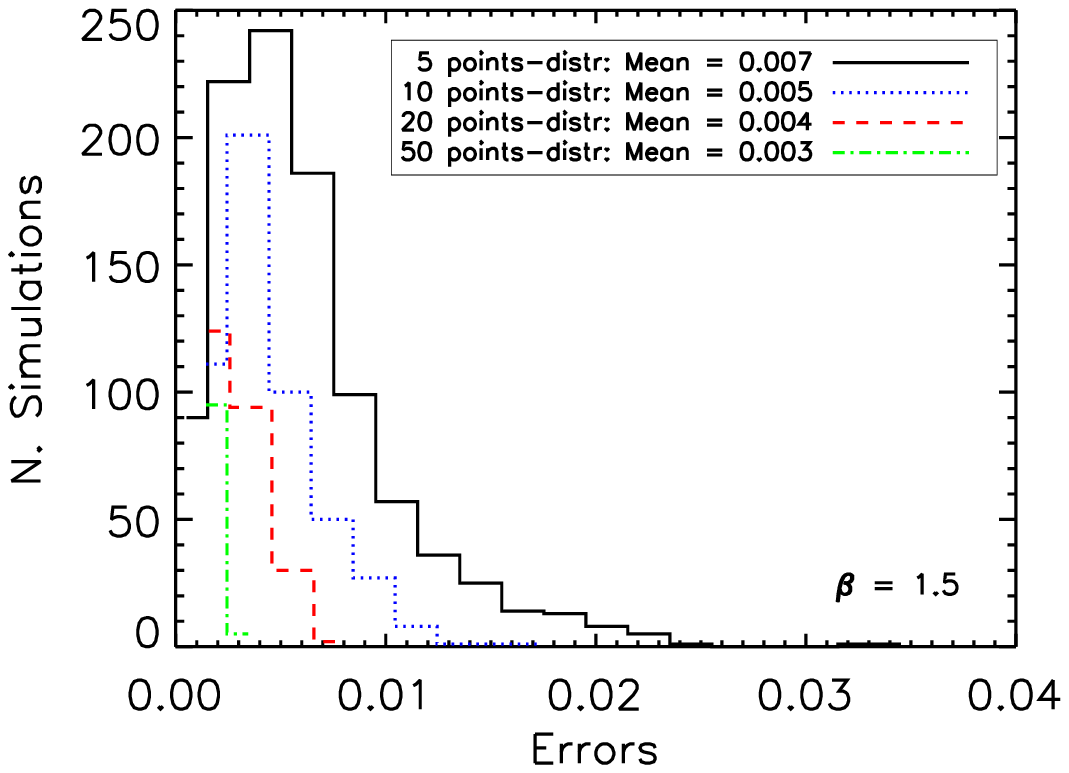} 
      \caption{As Figure \ref{fig:binningSN25} for a future mission with the characteristics and sampling pattern described in \S\ref{sec:7}, assuming a mcr of 0.1 cnt s$^{-1}$.}
\label{fig:histoWFXT}
\end{figure}

In order to verify the performance of such type of mission for faint AGN populations, 
we explored the dependence of the measured excess variance on different values 
of the source mean count rate. The results are summarized in Table \ref{tab:mcr_exv_bias}. 
The results are consistent with the findings shown in 
\S \ref{sec:5.1} (Table \ref{tab:snratio}) but now we are able to measure variability
with comparable accuracy at flux levels\footnote{Conversion factors from counts to fluxes 
were calculated assuming a power law spectrum with $\Gamma=1.4$ for
an unabsorbed AGN at $z=0$.} more than one order of magnitude lower than XMM,
using approximately the same observing time, thus allowing variability studies for hundreds
of AGN per square degree.
Such good performances are due in part to the larger effective area, and 
in part to the low background made possible by the considered low-earth orbital configuration.
We also notice that fortuitously, the bias actually improves at low count rates using such uniform sampling 
scheme, as the lost power 
better compensates the leakage coming from low frequencies.

\section{Summary and Conclusions}
\label{sec:8}

In this paper we studied the statistical properties of  
 \sNXV\ and its performance in the case of lightcurves with an intrinsic 
"red-noise" PSD. Strictly speaking, our results are valid in the case of Gaussian light 
curves, which remain only weakly non-stationary, on time scales of decades and shorter. Regarding the ``stationarity" of the variability 
process, this may not be such a restrictive assumption, if the emission mechanism(s) in most of the AGN we observe has reached  a 
``stable" state, i.e. statistical properties like the intrinsic mean flux, variance as well as the covariance at any two time points, may indeed 
remain constant, over time scales of the order of hundred of years or many decades. However, more work is necessary to investigate 
what will be the effects on our results if the AGN light curves are non-Gaussian.

Red-noise PSDs are common in many astrophysical time variable phenomena. 
In particular, the radio, optical and X-ray AGN lightcurves do show a "red-noise" behaviour. 
We performed detailed Monte Carlo simulations, assuming PSD slopes between 1 and 3, and we 
considered the case of continuously and sparsely sampled lightcurves, assuming various sampling patterns, and various S/N ratios. Our results can be summarized as follows: 

\begin{enumerate}
\item The statistical properties of  \sML\ and \sNXV\ are identical. Therefore, given the fact that it is easier to compute \sNXV, we propose its use for the study of the variability amplitude of the observed lightcurves. 
 
\item We study in detail the bias of the normalized excess variance. If the intrinsic mean of the lightcurve were known in advance, then \sNXV\ is an \textit{unbiased} estimate of the intrinsic source variance normalized to the mean squared, even in the case of sparsely sampled light curve. 
 
\item However, in most cases the intrinsic mean is unknown. In this case, our results show that \sNXV\ is a \textit{biased} estimate of \sbn\, even in the case of continuously sampled lightcurves. The bias depends on the PSD slope, $\beta$, and increases as the PSD slope steepens. As long as $\beta$ is known in advance, or ``red-noise'' PSD models are fitted to the observed \sNXV\ values, multiplication of the sampled \scon\ values by a factor equal to $0.48^{\beta-1}$ will result into estimates whose mean value will be within $\sim 15$\% of the intrinsic \sbn.  
 
\item The bias depends on the sampling pattern as well. However, even if there are many ``missing'' data points, the bias remains the same as in the case of the continuously sampled lightcurves, as long as the data are sampled uniformly over the observing period. 
In fact in general the sampling pattern is 
as important as, if not more important, than the total number of data points: the number
of effectively independent (clusters of) points is the primary factor affecting the estimate of the
total variance in cases where the variability amplitude is dominated by the longest timescales.
In the extreme sampling patterns (like for example when most of the data points were obtained at the start and at the end of the observing period) we suggest that the bias factor we mentioned above is multiplied by 1.3 in the case when $1\le \beta \le 2$. At steeper PSDs, the main factor that determines $b$ is the PSD slope and not the sampling pattern. 

\item Aliasing effects are negligible except for very flat PSD slopes with $\beta\lesssim 1$. Even for $\beta =1$
the correction is small and amounts to an increase of $5-10\%$ on the measured variance (and a corresponding decrease in the bias factor) depending on the sampling pattern. However aliasing is due to the PSD slope at the highest
sampled frequency and thus this additional correction is needed only for cases where the PSD is very flat or if we are only sampling long timescales below the break frequency.
 
\item Individual \sNXV\ measurements should be treated with extreme care, even if the ``bias correction'' recipe we mentioned above is applied to them. Their distribution has a large width, and is highly asymmetric, even in the case of continuously sampled lightcurves. The ratio of the distribution's width over its mean, and the skewness of the \sNXV\ distribution increase with $\beta$.  This is a well know result of \citet{Vaughan03}. However, we find that the confidence limits provided by these authors most probably do not apply in the case of lightcurves with a S/N ratio less than 3.  

\item For a given $\beta$,  the width and asymmetry of the \sNXV\ distribution increases significantly 
in the case lightcurves which are sparsely sampled, in a non-uniform pattern. Therefore, individual \sNXV\ measurements are even more unreliable in this case.  We do not recommend their use, specially in the case of extreme sampling patterns, and S/N ratios smaller than 3.

\item Based on our results, we strongly recommend the use of  ``ensemble''  \sNXV\ estimates in practice. These estimates should be preferred in the case when multiple lightcurves of the same object, or many lightcurves of objects with the same properties, are available.  If there are $n\ge 20$ of such lightcurves, the distribution of the \textit{mean}-\sNXV ~(as defined by Equation \ref{eq:ensemble_nxv}) will be quite symmetric (and well approximated by a Gaussian), and its standard deviation will be well approximated by the error of the \textit{mean}-\sNXV (given by Equation 13). This is result is valid for all PSD slopes, irrespective of the sampling pattern (i.e. whether the lightcurves are continuously or sparsely sampled), as long as the individual lightcurves have a S/N ratio $\ge 3$

\item At lower S/N ratio lightcurves, we recommend the use of the \textit{mean}-\sNXV, but with $n\ge 50$. 
\end{enumerate}

The normalized excess variance is a useful tool to characterize the variability amplitude of astrophysical sources in cases when the available lightcurves cannot be used to estimate the PSD of as source (i.e. it is short and/or has many gaps). We believe that our results will be helpful to future studies which employ X-ray \sNXV\ measurements to measure the BH mass of an AGN (both in the case of continuously and sparsely sampled lightcurves). The results presented in this work will be extremely useful in an era of rapidly growing (optical, radio, IR etc.) 
sky surveys, which often include timing informations despite the fact that a temporal strategy has not been 
specifically accounted for in planning the survey. In particular we showed that for a future X-ray mission, 
a properly designed observing strategy may allow to 
measure variability for hundreds of sources per square degree. 
Such dataset would largely overlap with the spectroscopic sample
\citep[e.g.][]{Gilli11}, thus resulting thousand of AGNs with both temporal and 
spectroscopic informations.
Since the individual variance estimates will still be affected by significant uncertainties,
a large dataset will be essential in order to constrain the average timing properties of high
redshift AGNs (provided that the AGN population shares the same intrinsic properties), and 
investigate their dependence of other source parameters (like spectral slope, luminosity etc). 
Several dedicated timing missions have also been proposed in the X-ray regime such as LOFT 
(Feroci et al. 2010). Our results are valid in such cases as well, as its instruments will provide 
data with  sampling patterns close to the continuous (large area monitor) or uniform 
(wide-field monitor) cases explored here. 
More in general, our results will apply to any source characterized by red-noise variability, and can thus be useful in the estimation of the variability amplitude of sources that will result from multi-epoch and time-domain surveys such as those provided by e.g. Pan-STARRS and LSST.

\section*{Acknowledgements}
MP acknowledges support from the Italian 
PRIN 2009 and FARO 2011 projects. Part of this work was supported by the COST 
Action MP0905 "Black Holes in a Violent Universe".

\label{lastpage}

\end{document}